\documentclass[10pt,pra,aps,showpacs,twocolumn,unsortedaddress]{revtex4-1}
\usepackage{float}
\usepackage{hyperref}
\usepackage{natbib}
\usepackage{amsmath}
\usepackage{amssymb, cancel}
\usepackage{verbatim}
\usepackage{bbold}
\usepackage{subfigure}

\usepackage{graphicx}

\DeclareSymbolFont{bbold}{U}{bbold}{m}{n}
\DeclareSymbolFontAlphabet{\mathbbold}{bbold}

\newcommand{\ve}{\varepsilon}

\newcommand{\bs}{\boldsymbol}

\newcommand{\oo}{\Omega}
\newcommand{\ua}{\uparrow}
\newcommand{\da}{\downarrow}
\newcommand{\tot}{\text{tot}}
\newcommand{\tsf}{\text{sf}}
\newcommand{\tnf}{\text{nf}}

\def\be#1\ee{\begin{equation}#1\end{equation}}
\def\ba#1\ea{\begin{align}#1\end{align}}
\def\bal#1\eal{\begin{equation}#1\end{equation}}

\begin{document}

\setlength{\abovedisplayskip}{4pt}
\setlength{\belowdisplayskip}{4pt}

\title{
Hydrodynamic modes of partially condensed Bose mixtures
}

\author{J. Armaitis}
\email{j.armaitis@uu.nl}
\author{H.T.C. Stoof}
\author{R.A. Duine}
\affiliation{Institute for Theoretical Physics and Center for Extreme
Matter and Emergent
Phenomena, 
Utrecht University, 
Leuvenlaan 4, 3584 CE Utrecht, The Netherlands}

\date{\today}

\begin{abstract}
We generalize the Landau-Khalatnikov hydrodynamic
theory for superfluid helium to
two-component (binary) Bose mixtures at arbitrary temperatures. In 
particular, we include the spin-drag terms
that correspond to viscous coupling between the clouds. 
Therefore, our theory not 
only describes the usual collective
modes of the individual components, e.g., first and second sound, but 
also results in new collective modes, where both constituents 
participate. We study these modes in detail and present their 
dispersions using thermodynamic quantities obtained 
within the Popov approximation.
\end{abstract}

% 03.75.Kk 	Dynamic properties of condensates; collective and hydrodynamic excitations, superfluid flow 
% 03.75.Mn - Multicomponent condensates; spinor condensates 
% 05.30.Jp 	Boson systems (for static and dynamic properties of Bose-Einstein condensates, see 03.75.Hh and 03.75.Kk; see also 67.10.Ba Boson degeneracy in quantum fluids)
\maketitle

\section{Introduction}

The realization of a Bose-Einstein condensate (BEC) in ultracold alkali-metal vapors \cite{BEC11,BEC12,BEC13} 
has ignited
a rapid progress in the understanding of degenerate
gases at low temperatures \cite{LeggettRev,RevManyBody}. A large part of this understanding 
has been gained through the study of collective modes \cite{StringariCol,CornellCol,KetterleCol}. In particular,
it has been shown that for a weakly interacting
gas of bosons close to absolute zero temperature, the
collective excitations are Bogoliubov quasiparticles \cite{BogTh58,GriffinTrap,AndersenRev,BogExp}
that are
responsible for fascinating properties of the system,
including superfluidity and quantum depletion of the condensate.

Work building upon the single-component ultracold gases has provided access to even richer
systems. In particular, considering 
mixtures of several species of particles with the same (Bose-Bose or
Fermi-Fermi mixtures  \cite{LiKmix, Fermi2ndSound13}) or different statistics (Bose-Fermi mixtures \cite{SolomonBFmix}) 
has become possible. 
These systems are known as binary mixtures or two-component gases.
Arguably the simplest of them is a mixture of two 
different hyperfine states of the same bosonic atom. However,
even this simple system poses important questions concerning the nature
of its ground state and the excitations. Therefore, much
work has been carried out on the static and dynamic properties
of the two-component Bose gas
 \cite{
%theory
ColsonFetter78,HoShenoy98,GrahamWalls98,EsryGreene99,
VortexCollective00,Timmermans03,DaltonGhanbari12,AbadRecati13,Zaremba13,PhysRevA.91.011602,
%experiment
CornellBinary98,InguscioSuper02,Hadzibabic13,EngelsPhaseWinding13}, both in the uniform case and 
for the trapped case. Most of the effort has been concentrated on 
the zero-temperature physics, with only a few studies \cite{Fertig07,Yu14} on the 
properties of binary Bose mixtures at nonzero temperature.

Having more than a single component in the gas also allows one to make
a connection to the physics of spins, by introducing a
pseudospin to distinguish
the two components. In particular, one can consider
ferromagnetic and antiferromagnetic states \cite{Ho98,Machida98,StamperKurnUeda13} 
as well as spin dynamics \cite{Chapman05,StamperKurn06},
and topological spin textures
\cite{DiracMonopoles, AlKhawaja2001Nat, AlKhawaja2001PRL, Ho98, Machida98, BigelowTopo, SkyrmionsExp,ArmaitisPRL1,PRL1Exp, UedaKnots}.
One kinetic effect concerning the spin dynamics is
the so-called spin drag \cite{PhysRevLett.98.266403,spindragTh}. 
This recently observed \cite{Sommer2011,spindragExp} effect corresponds to the relaxation of the 
difference of the velocities between the two components. Understanding the 
interplay between the BEC, the thermal particles,
and the spin degrees of freedom in this relatively simple
and well-controlled two-component Bose gas might also offer some 
insights for interacting spinful degenerate systems of a rather
different nature, such as the condensate of magnons \cite{MagnonsExp,MagnonsTh}.

In this paper we build upon our previous results for the ferromagnetic
Bose gas \cite{ArmaitisPRL1}, but now consider a different situation where two condensates 
are present in the miscible regime. We tackle the problem of the 
collective modes of the two-component mixture both in the uniform
gas and in a trap in an effort toward making a connection with experiments. The
structure of the paper is as follows. 
In Sec.\ II we describe the Popov theory of the 
binary Bose mixture, and present relevant thermodynamic functions in that
approximation, including the equation of state. 
We study the effects of spin drag in Sec.\ III.
In Sec.\ IV we develop
a linear hydrodynamic model that makes use of our
previous thermodynamic results and describes
a two-component system accounting for spin drag. We present the results for the uniform and trapped gas in Sec.\ V. Finally, we
conclude in Sec.\ VI.

\section{Microscopic theory}

In this section we briefly describe the microscopic Popov theory of the
two-component Bose mixture. The Popov theory is an extension of the
Bogoliubov theory to relatively high temperatures, which includes an improved treatment of the excitations. Specifically,
the Bogoliubov excitations are allowed to interact, and their interactions are treated
in the Hartree-Fock approximation. Multi-component gases of bosons have been
treated in the Bogoliubov framework before \cite{Bassichis64}. In 
particular, two-component mixtures have been considered
in Refs.\ \cite{Larsen63, Timmermans03, Zaremba13}, and some results from the
Popov theory have been presented in Ref.\ \cite{Yu14}.
The novelty of our results is twofold: We present
the Popov analysis in the functional-integral formalism
and calculate the thermodynamic properties of the balanced 
binary Bose gas.
Our discussion on the Bogoliubov transformation follows the usual
grand-canonical treatment of the problem. The
single-component situation has been treated 
in this way in, for instance,  Refs.\ \cite{Falco07,stoofbook}.

\subsection{General binary mixture}

In general, a grand-canonical partition function 
for two bosonic fields ($\phi_\ua$ and $\phi_\da$) that includes all the possible $s$-wave
interactions is
\be
Z = \int d[\phi_\ua^*] d[\phi_\ua] d[\phi_\da^*] d[\phi_\da]
e^{-S[\phi_\ua^*,\phi_\ua,\phi_\da^*,\phi_\da]/\hbar},
\ee
where the action is
%
%
%\begin{widetext}
\ba
S&[\phi_\ua^*,\phi_\ua,\phi_\da^*,\phi_\da]
=
\int_0^{\hbar\beta} d \tau \int d\bs x
\\ \nonumber 
&\times
\Bigg(
\sum_{\sigma=\ua,\da}
\Bigg[
\phi_\sigma^*
\bigg(
\hbar \partial_\tau
-
\frac{\hbar^2 \bs \nabla^2}{2m_\sigma}
-
\mu_\sigma
\bigg)
\phi_\sigma
\nonumber\\ \nonumber
&
+\frac{1}{2}
g_{\sigma\sigma}
\phi_\sigma^* \phi_\sigma^* \phi_\sigma \phi_\sigma
\Bigg]
+
g_{\ua\da}
\phi_\ua^* \phi_\da^* \phi_\da \phi_\ua
\Bigg)
,
\ea
%\end{widetext}
and all the fields are considered at the position $\bs x$ and
the imaginary time $\tau$. Moreover, $\beta=1/k_\text{B}T$ is the inverse thermal energy, $m_\sigma$ are the masses of the particles,
and $g_{\sigma\sigma'}$ are the two-body $T$ matrices describing
the $s$-wave interactions.

From now on we focus on the symmetric case, where 
the masses are equal $m=m_\ua=m_\da$, and the intraspecies interactions
are the same and described by a single scattering length $a$. Thus,
\be
g=g_{\ua\ua}=g_{\da\da}=\frac{4\pi\hbar^2 a}{m}.
\ee
The interspecies interactions are described by another
scattering length $a_{\ua\da}$, implying that
\be
g_{\ua\da}=\frac{4\pi\hbar^2 a_{\ua\da}}{m}.
\ee
All the interactions are assumed to be repulsive, i.e., $g>0$ and
$g_{\ua\da}>0$.
Furthermore, as opposed to our earlier work in Ref.\ \cite{ArmaitisPRL1},
we here focus on the case with two separate condensates. In order
for this to be possible, the condition
\be
g > g_{\ua\da}
\ee
has to be satisfied, as otherwise the two components demix \cite{PandSbook}.
%pg 351, eq 12.14

We are now in a position to perform a fluctuation expansion
for each species by putting
\be
\phi_\sigma(\bs x, \tau) = \phi_{0\sigma}(\bs x)
+ \phi_{\sigma}'(\bs x, \tau),
\ee
where the fluctuations $\phi_{\sigma}'(\bs x, \tau)$ 
are, on average, zero. Moreover, the fluctuations are
orthogonal to the condensate 
$\langle \phi_\sigma(\bs x, \tau) \rangle = \phi_{0\sigma}(\bs x)$
of the same species which means that
\be
\int d \bs x \bigg(
\phi_{0\sigma}^*(\bs x)
\phi_{\sigma}'(\bs x, \tau)
+
\phi_{0\sigma}(\bs x)
\phi_{\sigma}'^*(\bs x, \tau)
\bigg)
=0.
\ee
Since in what follows the relative phases of the condensates
do not play a significant role, we choose both the condensate fields to be real,
\be
\phi_{0\sigma}(\bs x) = \sqrt{n_{0\sigma}},
\ee
where $n_{0\sigma}$ is the atomic (number) density of the condensed
$\sigma$ particles. Moreover, since we are considering the
uniform case here, the condensate density has no spatial
dependence. Expanding the fields in the action in this manner,
we have for the action
\be
S = 
S_0 +
\sum_{\sigma=\ua,\da} S_{1\sigma} + 
S_2
+
S_3
+
S_4,
\ee
where the zeroth-order (Landau-free-energy) contribution is
\ba
S_0
=
\hbar\beta V 
\big(
&-\mu_\ua n_{0\ua}
+
g n_{0\ua}^2 / 2
\nonumber \\
&-\mu_\da n_{0\da}
+
g n_{0\da}^2 / 2
+
g_{\ua\da} n_{0\ua} n_{0\da}
\big)
,
\ea
the term linear in fluctuations reads
\ba
S_{1\sigma}[\phi_{\sigma}'^*,\phi_{\sigma}']
=
\int_0^{\hbar\beta} d \tau \int d\bs x
\,
\Bigg[
\phi_{\sigma}'^*
\big(
-
\mu_\sigma
+ g n_{0\sigma}
\nonumber \\
+ g_{\ua\da} n_{0\bar \sigma}
\big)
\sqrt{n_{0\sigma}}
+ \text{c.c.} \Bigg],
\ea
and the quadratic term is
\begin{widetext}
\ba
S_2[\phi_{\ua}'^*,\phi'_\ua,\phi_{\da}'^*,\phi'_\da]
&=
\int_0^{\hbar\beta} d \tau \int d\bs x
\, %\bigg\{
\sum_{\sigma=\ua,\da}
\bigg[
\phi_{\sigma}'^*
\bigg(
\hbar \partial_\tau
-
\frac{\hbar^2 \bs \nabla^2}{2m}
-
\mu_\sigma
+ 2 g n_{0\sigma}
+ g_{\ua\da} n_{0\bar \sigma}
\bigg)
\phi_{\sigma}'
\nonumber \\
&
+
\frac{g n_{0\sigma}}{2}
\big(
\phi_{\sigma}'^*\phi_{\sigma}'^*
+
\phi_{\sigma}'\phi_{\sigma}'
\big)
\Bigg]
+
g_{\ua\da}\sqrt{n_{0\ua}n_{0\da}}
\int_0^{\hbar\beta} d \tau \int d\bs x
\big(
\phi_{\ua}'\phi_{\da}'
+\phi_{\ua}'^*\phi_{\da}'
+\phi_{\ua}'\phi_{\da}'^*
+\phi_{\ua}'^*\phi_{\da}'^*
\big),
\ea
\end{widetext}
where all the fluctuation fields are evaluated at $(\bs x, \tau)$, and we have denoted the species opposite to $\sigma$ by $\bar \sigma$.
Furthermore, $S_3$ and $S_4$ terms describe the interactions between the fluctuations,
\begin{widetext}
\ba
S_3[\phi_{\ua}'^*,\phi'_\ua,\phi_{\da}'^*,\phi'_\da]
=
\int_0^{\hbar\beta} d \tau \int d\bs x
\, 
%\bigg[
&\bigg(\sum_{\sigma=\ua,\da}
g\sqrt{n_{0\sigma}} 
\left(
\phi_{\sigma}'^*\phi_{\sigma}'^*\phi_{\sigma}' + 
\phi_{\sigma}'\phi_{\sigma}'\phi_{\sigma}'^*
\right)
\nonumber \\
&+
g_{\ua\da}
\sqrt{n_{0\ua}}
(
\phi_{\da}'^*\phi_{\da}'\phi_{\ua}' +
\phi_{\da}'\phi_{\da}'^*\phi_{\ua}'^*
)
+
g_{\ua\da}
\sqrt{n_{0\da}}
(
\phi_{\ua}'^*\phi_{\da}'\phi_{\ua}'
+
\phi_{\ua}'\phi_{\da}'^*\phi_{\ua}'^*
)
\bigg)
,
\ea
\end{widetext}
and
\ba
S_4[\phi_{\ua}'^*,\phi'_\ua,\phi_{\da}'^*,\phi'_\da]
&=
\\
\int_0^{\hbar\beta} d \tau \int 
d\bs x
\, %\bigg\{
&\bigg(\sum_{\sigma=\ua,\da} 
\frac{g}{2}
\phi_{\sigma}'^*\phi_{\sigma}'^*\phi_{\sigma}' \phi_{\sigma}'
+
g_{\ua\da}
\phi_{\ua}'^*\phi_{\da}'^*\phi_{\da}' \phi_{\ua}'
\bigg)
.
\nonumber 
\ea
Note that the terms $S_3$ and $S_4$ are neglected in the Bogoliubov
theory.

We now perform the Hartree-Fock theory for the excitations, which 
involves the inclusion of the mean field,
\be
\langle \phi_{\sigma}'^*\phi_{\sigma}' \rangle
= n'_\sigma,
\ee
where $n'_\sigma$ is the density of the excitations of
the $\sigma$ species such that the total density of a species is
\be
n_\sigma = n_{0\sigma} + n'_\sigma.
\ee
Note that we neglect the
coherence between the two species and take $\langle \phi_{\sigma}'^*\phi_{\bar \sigma}' \rangle=0$. Therefore, the appropriate mean-field substitutions are
\ba
\phi_{\sigma}'^*\phi_{\sigma}'^*\phi_{\sigma}' \phi_{\sigma}'
\rightarrow
4n'_\sigma \phi_{\sigma}'^*\phi_{\sigma}'
-
2{n'_\sigma}^2,
\\
\phi_{\ua}'^*\phi_{\da}'^*\phi_{\da}' \phi_{\ua}'
\rightarrow
n'_\ua \phi_{\da}'^*\phi_{\da}'
+
n'_\da \phi_{\ua}'^*\phi_{\ua}'
-
n'_\da n'_\ua,
\ea
where the subtractions account for double counting
in the quartic term of the action, whereas
in the cubic terms no double-counting problems appear, as can be seen by applying
Wick's theorem.

By requiring all the linear terms in $\phi'_\sigma$ and $\phi_\sigma'^*$
of the action to vanish,
we obtain a set of two Gross-Pitaevskii equations for the uniform
condensates that read
\ba
(-\mu_\ua + g n_{0\ua} + g_{\ua\da} n_{0\da}
+ 2g n'_\ua
+ g_{\ua\da} n'_\da
) 
\sqrt{n_{0\ua}}
=0,\\
(
-\mu_{\da}
+g n_{0\da}
+g_{\ua\da} n_{0\ua}
+2 g n'_\da
+g_{\ua\da}  n'_\ua
)
\sqrt{n_{0\da}}
=0,
\ea
and from which the chemical potentials are obtained as
\ba
\mu_\ua = g n_{0\ua} + g_{\ua\da} n_{0\da}
+ 2g n'_\ua
+ g_{\ua\da} n'_\da,\\
\mu_{\da}=
g n_{0\da}
+g_{\ua\da} n_{0\ua}
+2 g n'_\da
+g_{\ua\da}  n'_\ua.
\label{eq:chempots}
\ea
In order to diagonalize the quadratic part of the action, we perform a Fourier transformation, and then introduce 
Nambu space \cite{stoofbook}. Since we want to rewrite the quadratic part of the 
action in the form
\be
S_2[\phi_{\ua}'^*,\phi'_\ua,\phi_{\da}'^*,\phi'_\da] 
= 
-\frac{\hbar}{2}
\sum_{\bs k\neq 0, n} \bs \Phi_{\bs kn} \cdot \bs G^{-1}_{\bs kn} \cdot {\bs \Phi}_{\bs kn}^\dagger,
\ee
where $\hbar \bs k$ is the momentum, $n$ labels the
Matsubara frequencies $\omega_n = 2 \pi n / \hbar \beta$,
\be
\bs \Phi_{\bs kn} = \big(\phi_{\ua \bs k n}'^*, \phi_{\ua -\bs k n}',
\phi_{\da \bs k n}'^*,\phi_{\da -\bs k n}'\big),
\ee
is a vector in the appropriate Nambu space in this case, 
and $\bs G^{-1}$ is the inverse Green's function of the system. Note that we have to take care to preserve the correct time ordering. The latter results in an extra term in the action,
\ba
S_\text{TO}
=
-\frac{\hbar\beta}{2}
\sum_{\sigma,\bs k\neq 0, n}
\big[
\ve_{\bs k}
-
\mu_\sigma
+ g (2n_{0\sigma}+2n'_\sigma)
\nonumber\\
+ g_{\ua\da} (n_{0\bar \sigma}+n'_{\bar \sigma})
\big]
=
-\frac{\hbar\beta}{2}
\sum_{\sigma,\bs k\neq 0, n}
\left(
\ve_{\bs k}
+ g n_{0\sigma}
\right),
\ea
where $\ve_{\bs k} = \hbar^2 k^2/2m$ is the kinetic energy. Therefore,
plugging the expressions for the chemical potentials $\mu_\sigma$ into the action, we have
\ba
S =
&-\hbar\beta V
\bigg(
g_{\ua\da} n_\ua n_\da
+
g
(n_\da^2+n_\ua^2)
 \\ \nonumber
&-
g\frac{
n_{0\da}^2 
+
n_{0\ua}^2
}{2}
\bigg)
-\frac{\hbar\beta}{2}
\sum_{\sigma,\bs k\neq 0}
\left(
\ve_{\bs k}
+ g n_{0\sigma}
\right)
 \\ \nonumber
&-
\frac{\hbar}{2}
\sum_{\bs k\neq 0, n} \bs \Phi_{\bs kn} \cdot \bs G^{-1}_{\bs kn} \cdot {\bs \Phi}^\dagger_{\bs kn},
\ea
where $V$ is the volume of the system, while
\ba
-\hbar \bs G^{-1}_{\bs kn}
=
 \begin{pmatrix}
-\hbar G^{-1}_{\text B, \ua\bs kn} & \hbar \Sigma_{\ua\da} \\
\hbar \Sigma_{\ua\da} & -\hbar G^{-1}_{\text B, \da\bs kn}
 \end{pmatrix},
\ea
where $G^{-1}_{\text B, \ua\bs kn}$ and $\Sigma_{\ua\da}$ are two-by-two submatrices (from now on
two-by-two matrices are denoted by capital letters, while
four-by-four matrices are denoted by bold capital letters). 
The submatrices on the diagonal are exactly the same as the Bogoliubov (single-component) inverse Green's functions,
that is,
\ba
-\hbar G^{-1}_{\text B, \sigma\bs kn}
=
 \begin{pmatrix}
 -i\hbar\omega_n +\ve_{\bs k} + gn_{0\sigma} & gn_{0\sigma} \\
  gn_{0\sigma} & i\hbar\omega_n +\ve_{\bs k} + gn_{0\sigma}
 \end{pmatrix},
\ea
and the off-diagonal matrix is the self-energy due to the interspecies coupling given by
\ba
\hbar \Sigma_{\ua\da}
= g_{\ua\da} \sqrt{n_{0\ua} n_{0\da}}
 \begin{pmatrix}
  1 & 1 \\
  1 & 1
 \end{pmatrix}.
\ea

Anticipating the Bogoliubov transformation we define
\ba
\bs I = 
 \begin{pmatrix}
  \sigma_z & 0 \\
  0 & \sigma_z
 \end{pmatrix},
\ea
where
\ba
\sigma_z = 
 \begin{pmatrix}
  1 & 0 \\
  0 & -1
 \end{pmatrix}.
\ea
Moreover, we define a matrix $\bs \Gamma_{\bs k}$ which is obtained by setting 
$\omega_n$ to zero in the inverse Green's function, i.e.,
\be
\bs \Gamma_{\bs k} = -\hbar \bs G^{-1}_{\bs k 0}.
\ee
We now proceed to diagonalize the matrix $\bs \Gamma_{\bs k}$ while keeping the bosonic
character of the excitations, i.e., requiring that
their operators obey bosonic 
commutation relations. In our formalism the preservation
of the bosonic commutation relations is 
represented by the fact that the Matsubara-frequency terms
are unaffected by the transformation. Hence, we
look for real eigenvectors $\bs w$ of $\bs I \cdot \bs \Gamma_{\bs k}$,
\be
\bs I \cdot \bs \Gamma_{\bs k} \cdot \bs w_{\pm,\bs k} = E_{\pm,\bs k} \bs w_{\pm,\bs k},
\ee
satisfying the property
\be
\bs w_{\pm,\bs k} \cdot \bs I \cdot \bs w_{\pm,\bs k} = 1.
\ee
The above-mentioned eigenvalues define the dispersions
of the Bogoliubov quasiparticles 
\ba
E_{\pm,\bs k}^2
=
&\ve_{\bs k}(\ve_{\bs k}+ g(n_{0\ua}+n_{0\da}))
\\ \nonumber
&\pm
\ve_{\bs k}
\sqrt{
g^2 (n_{0\ua} - n_{0\da})^2 + 4 g_{\ua\da}^2 n_{0\ua}n_{0\da} 
}.
\ea
The eigenvectors $\bs w_{\pm,\bs k}$ have the so-called Bogoliubov
coherence factors as their entries, or explicitly
\be
\bs w_{\pm,\bs k}
=
(u_{\pm,\bs k}^\ua, -v_{\pm,\bs k}^\ua, u_{\pm,\bs k}^\da, -v_{\pm,\bs k}^\da).
\ee
For the case at hand these coherence factors are rather involved
functions of the momentum. Since they have been presented explicitly
in Ref.\ \cite{Zaremba13}, we do not write them out here. 

The relation between the two original fluctuation fields 
and the new Bogoliubov quasiparticle fields is given by
\ba
\bs \Phi_{\bs kn} = \bs W_{\bs k} \cdot \bs \Psi_{\bs kn},
\ea
where we have defined the Bogoliubov fields
\be
\bs \Psi_{\bs kn} = \big(\psi_{+, \bs k n}^*, \psi_{+, -\bs k n},
\psi_{-, \bs k n}^*,\psi_{-, -\bs k n}\big),
\ee
and the transformation matrix
\ba
\bs W_{\bs k}
=
 \begin{pmatrix}
W^\ua_{+,\bs k} & W^\ua_{-,\bs k} \\
W^\da_{+,\bs k} & W^\da_{-,\bs k}
 \end{pmatrix},
\ea
which consists of the submatrices
\ba
W^\sigma_{s,\bs k}
=
 \begin{pmatrix}
 u^\sigma_{s,\bs k} & -v^\sigma_{s,\bs k} \\
  -v^\sigma_{s,\bs k} & u^\sigma_{s,\bs k}
 \end{pmatrix}.
\ea
It is straightforward to check
that this transformation leaves the Matsubara-frequency terms
in the action unaffected, and therefore the Bogoliubov excitations
are bosons. The action becomes
\ba
S =
&-\hbar\beta V
\left(
g_{\ua\da} n_\ua n_\da
+
g
(n_\da^2+n_\ua^2)
-
g\frac{
n_{0\da}^2 
+
n_{0\ua}^2 
}{2}
\right)
\nonumber \\
&+\frac{\hbar\beta}{2}
\sum_{\bs k\neq 0}
\left(
E_{+,\bs k} + E_{-,\bs k}
-2\ve_{\bs k}
- g (n_{0\ua}+n_{0\da})
\right)
\nonumber \\
&+
\sum_{s=\pm}
\sum_{\bs k\neq 0, n} 
(
-i\hbar\omega_n + E_{s,\bs k}
)
\psi_{s, \bs k n}^*, \psi_{s, \bs k n},
\ea
where $(E_{+,\bs k} + E_{-,\bs k})$ in the second term is again due to the time ordering but this time of the $\psi_\pm$ fields.

In the preceding equation, the first term describes the Hartree-Fock
contribution to the action. The second term (after 
properly accounting for
the fact that the contact potential does not fall off
at high momenta) can be shown to describe the so-called
Lee-Huang-Yang correction \cite{Lee57-1,Lee57-2},
which is small for a weakly interacting gas, and can therefore
be safely neglected. The last term describes the Bogoliubov
excitations. Evaluating this path integral amounts to a Gaussian
integration and can be performed exactly. Finally, we
perform the remaining bosonic Matsubara sum \cite{stoofbook}
\ba
\lim_{\eta \rightarrow 0^+}
\sum_n
\ln\left[\beta(-i\hbar\omega_n + E_{s,\bs k})\right]e^{i\omega_n \eta}
\nonumber \\
= \ln(1-e^{-\beta E_{s,\bs k}}),
\ea
and arrive at the following expression for the partition function
\ba
Z = \exp\bigg[
\beta
\big(
g_{\ua\da} n_\ua n_\da
+
g
(n_\da^2+n_\ua^2)
-
g\frac{
n_{0\da}^2 
+
n_{0\ua}^2 
}{2}
\big)
\nonumber \\
-
\frac{1}{V}\sum_{s=\pm,\bs k \neq 0} \ln(1-e^{-\beta E_{s,\bs k}})
\bigg].
\ea

The partition function is related to the pressure by
\ba
p(n_{0\ua}, n_{0\da}, T) &= -\Omega/V = \frac{1}{\beta}\ln(Z)
\nonumber \\
&=
%\left(
g_{\ua\da} n_\ua n_\da
+
g
(n_\da^2+n_\ua^2)
%\right)
-
g\frac{
n_{0\da}^2 
+
n_{0\ua}^2 
}{2}
\nonumber \\
&\phantom{=}
-
\frac{1}{\beta V}\sum_{s=\pm,\bs k \neq 0} \ln(1-e^{-\beta E_{s,\bs k}}),
\label{eq:pressure}
\ea
where $\Omega$ is the grand potential. We have obtained the (average) particle densities
\be
n_\sigma=
n_{\sigma0} +
\frac{1}{V}\sum_{{s}=\pm,\bs k \neq 0} \frac{|u_{s,\bs k}^\sigma|^2+|v_{s,\bs k}^\sigma|^2}{e^{\beta E_{s}}-1}
\ee
from the appropriate Green's functions, while the entropy per volume \cite{omega} is
\ba
\frac{S}{V}
=
-
\frac{\partial \oo}{\partial T}
{\bigg \vert}_{{n'_\sigma},{n_{0\sigma}}}
=
&-\frac{k_B}{V}\sum_{s=\pm,\bs k \neq 0} \ln(1-e^{-\beta E_{s,\bs k}})
\nonumber \\
&+
\frac{k_B}{V}\sum_{s=\pm,\bs k \neq 0} \frac{\beta E_{s,\bs k}}{e^{\beta E_{s,\bs k}}-1}.
\ea
For future convenience, we define the entropy per particle as
\be
s \equiv \frac{S}{V}\frac{1}{n_\ua + n_\da}
\ee
and also the total chemical potential $\mu_\tot$,
as well as the difference of the chemical potentials
\ba
\mu_\tot & = \mu_\ua + \mu_\da, \\
\Delta \mu & = \mu_\ua - \mu_\da.
\ea
Similarly, we define the total particle density and the difference of the particle
densities:
\ba
n_{\tot} & = n_{\ua} + n_{\da}, \\
\Delta n & = n_{\ua} - n_{\da}.
\ea
Finally, it is also beneficial to define two additional $T$ matrices,
\be
g_\pm = g \pm g_{\ua\da} = g (1 \pm \gamma),
\ee
where we have also introduced a dimensionless number $\gamma$, which
shows the relative strength between the interspecies  and the
intraspecies repulsion. Note that in the miscible case that we
discuss here, $\gamma < 1$.

\subsection{Balanced mixture}

Of particular interest is the balanced case, where the number of particles of the two species are equal: $n_\ua=n_\da=n$. We consider it
in this subsection. An obvious consequence of this limit is
\be
n_{0\ua} = n_{0\da} \equiv n_0,
\ee
and
\be
\mu \equiv \mu_\ua = \mu_\da = g_+ n + g n'.
\ee
Moreover, the dispersions of the quasiparticles become
\be
E_{\pm,\bs k}^2 = \ve_{\bs k} (\ve_{\bs k} + 2 n_0 g_\pm),
\label{eq:balbog}
\ee
and the Bogoliubov transformation matrix simplifies considerably to
\ba
\bs W_{\bs k}
=
 \begin{pmatrix}
 W_{+,\bs k} & W_{-,\bs k} \\
  W_{+,\bs k} & W_{-,\bs k}
 \end{pmatrix},
\ea
which can be inverted to 
\ba
\bs W_{\bs k}^{-1}
= \frac{1}{2}
 \begin{pmatrix}
 W_{+,\bs k}^{-1} & W_{+,\bs k}^{-1} \\
  -W_{-,\bs k}^{-1} & W_{-,\bs k}^{-1}
 \end{pmatrix},
\ea
from which we can conclude that the $\psi_{+,\bs kn}$ field
has equal contributions from the $\phi_{\ua\bs k n}$ and $\phi_{\da\bs kn}$
fields, whereas the latter fields enter $\psi_{-,\bs kn}$
with a relative minus sign but with equal absolute weights. This 
implies that $\psi_{+,\bs kn}$
describes densitylike excitations, whereas $\psi_{-,\bs k}$ describes
spinlike excitations.
The submatrices $W_{s,\bs k}$ in the above are
\ba
W_{s,\bs k}
=
 \begin{pmatrix}
 u_{s,\bs k} & -v_{s,\bs k} \\
  -v_{s,\bs k} & u_{s,\bs k}
 \end{pmatrix},
\ea
where in this case the coherence factors are simple
enough to be written out explicitly as
\ba
v_{\pm,\bs k}^2 =\frac{1}{4}\left(\frac{\ve_p + g_\pm n_0}{E_{\pm,\bs k}}-1\right),\\
u_{\pm,\bs k}^2 =\frac{1}{4}\left(\frac{\ve_p + g_\pm n_0}{E_{\pm,\bs k}}+1\right).
\ea
Note that the coherence factors are very similar to the
single-species case. However, the prefactor 
($1/4$) here is different from the single-species case ($1/2$),
since the transformation now involves four fields instead of two.

We now proceed to discuss the thermodynamic functions of the 
balanced binary mixture. Throughout the discussion, we consider
three different dimensionless interaction parameters 
$n^{1/3} a = 0.01$, $0.05$, and $0.1$. They have been chosen
to correspond to the experimentally relevant weakly interacting
(far away from Feshbach resonances) ultracold gas situations.
In particular, we consider the sodium atom which has 
several scattering lengths between the accessible hyperfine
levels close
to $50$ Bohr radii \cite{NaScattering}. Moreover, we are
interested in the hydrodynamic regime, where the
density in the center of the trap might become as high
as $10^{21}$~m$^{-3}$ \cite{PeterHeat}, which corresponds
to $n^{1/3} a \simeq 0.03$. Comparing thermodynamic quantities 
calculated within the
Bogoliubov theory with the renormalization group results
for the single-species case (see Ref.\ \cite{RG})
shows that the two agree very well all the way up to the condensation
temperature $T_c$ for $n^{1/3} a = 0.01$. Agreement for 
$n^{1/3} a = 0.05$ and particularly for $0.1$ is less good. However, 
the 
qualitative features of the thermodynamic functions are preserved. 
We expect the situation to be similar for the two-species case.

\begin{figure*}[htp]
%1125-2species-pressure
\centering
\includegraphics[width=.3\textwidth]{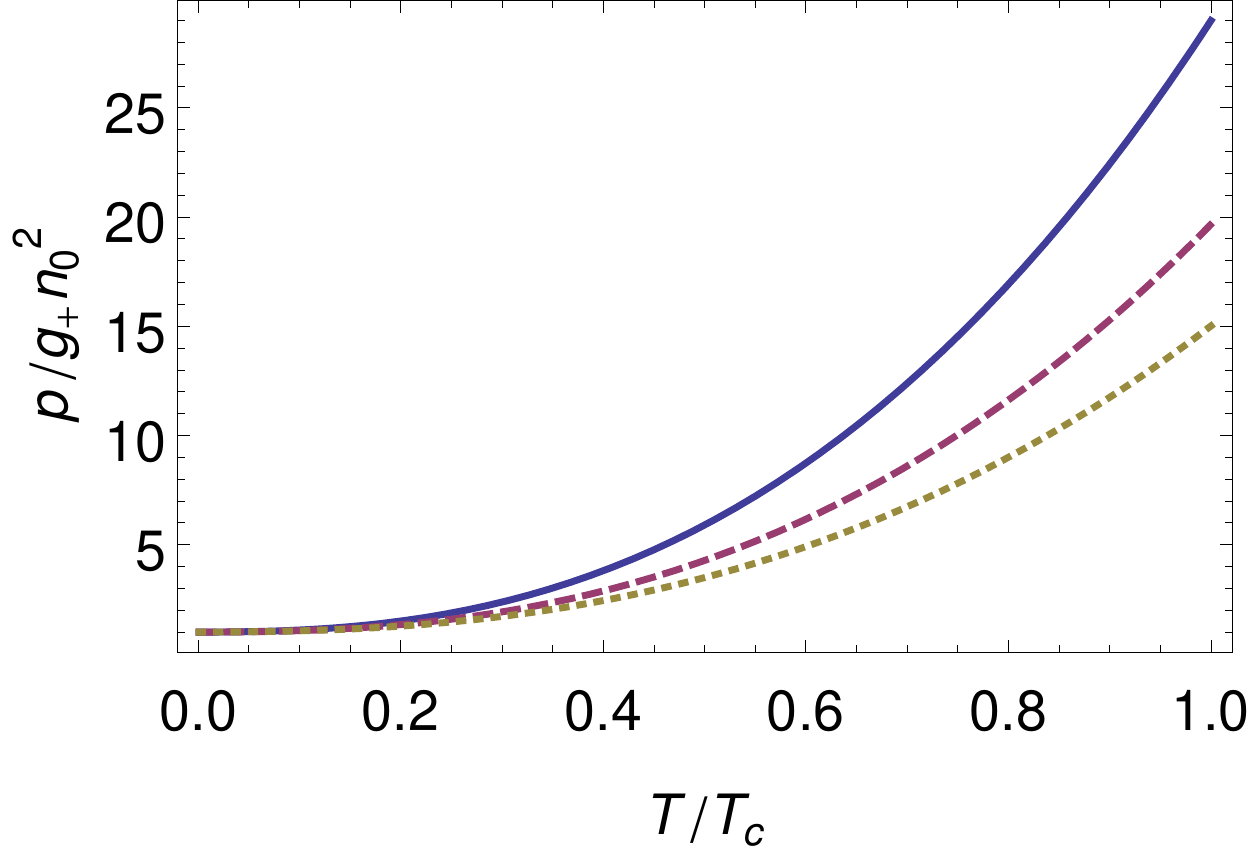}
\includegraphics[width=.3\textwidth]{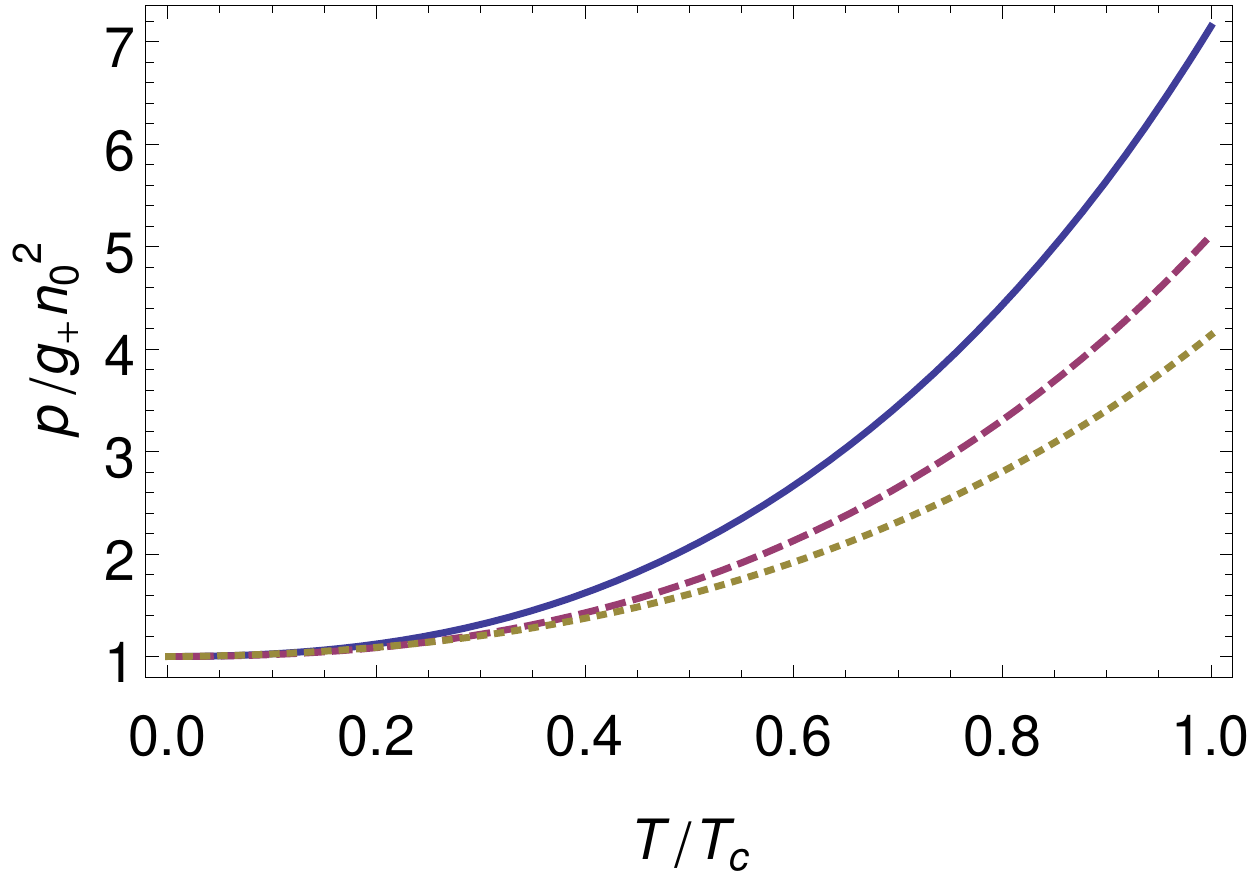}
\includegraphics[width=.3\textwidth]{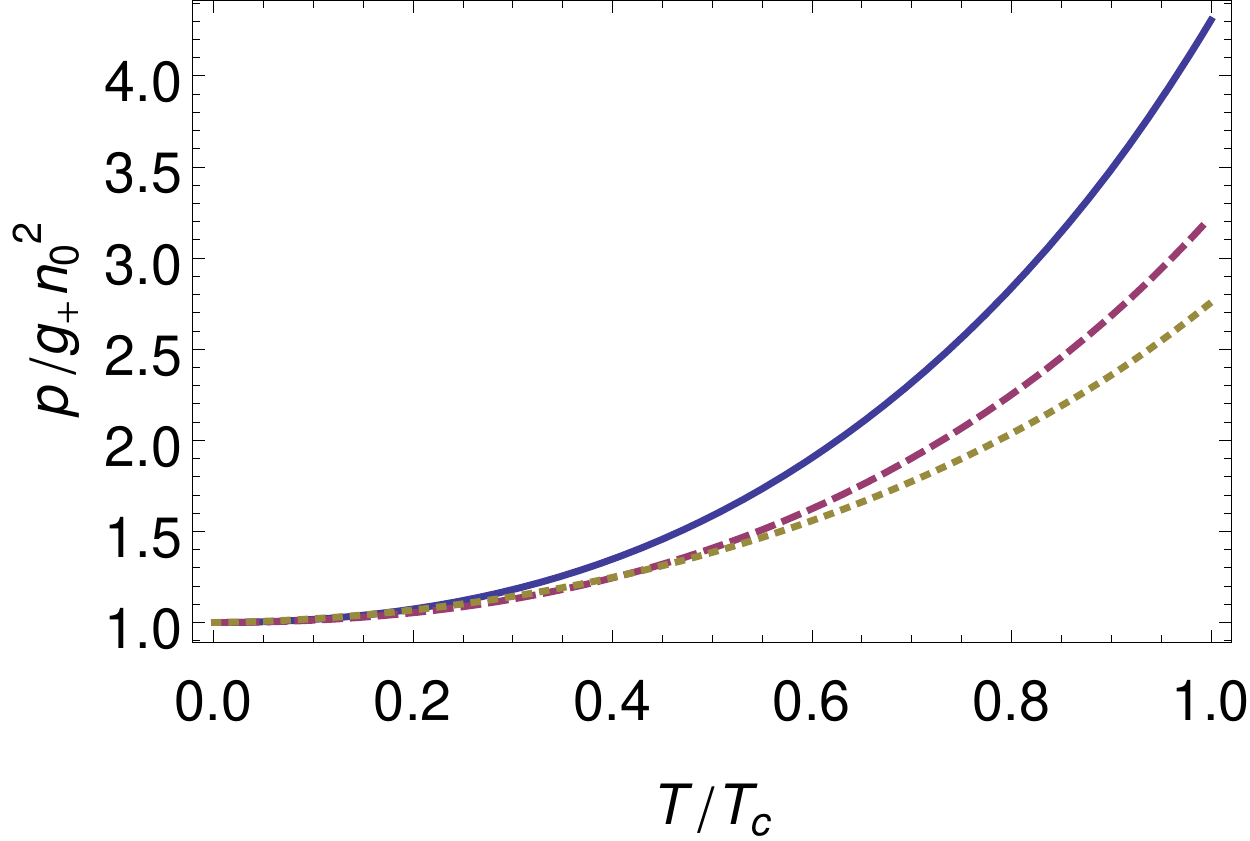}
\caption{(Color online) Pressure of the two-species balanced gas.
The interaction parameter increases from left to right: $n^{1/3} a = 0.01$, $0.05$, and $0.1$. Intercomponent interactions are
zero ($\gamma=0$) for the solid line, moderate ($\gamma=1/2$)
for the dashed line, and strong ($\gamma\lesssim 1$) for the dotted line.}
\label{fig-pressure}
\end{figure*}

We present the equation of state (pressure) in
Fig.\ \ref{fig-pressure}. We scale
the pressure plots with the zero-temperature pressure $g_+ n^2/2$
obtained from a Gross-Pitaevskii calculation \cite{PandSbook}.
For increasing interaction strength, the scaled pressure
decreases, since the thermal energy becomes comparable with
the interaction energy at higher temperature.
For $\gamma = 0$, the pressure is equal to the pressure of a single
species of gas with twice the density (see, e.g., Ref.\ \cite{RG} for
comparison).

Note that for a fixed total number of particles, the pressure is 
not a monotonically increasing function
of temperature in the Bogoliubov theory, as opposed
to the Popov theory. This spurious effect appears
due to the competition between the depletion of the condensate
(decreases pressure) and the population of the 
thermal states (increases pressure). Moreover, this leads
to a spurious lack of avoided crossing of the first- and second-sound velocities, therefore necessitating the use of at least
the Popov theory to describe the sound velocities accurately. 

\begin{figure*}[htp]
%1125-2species-pressure
\centering
\includegraphics[width=.3\textwidth]{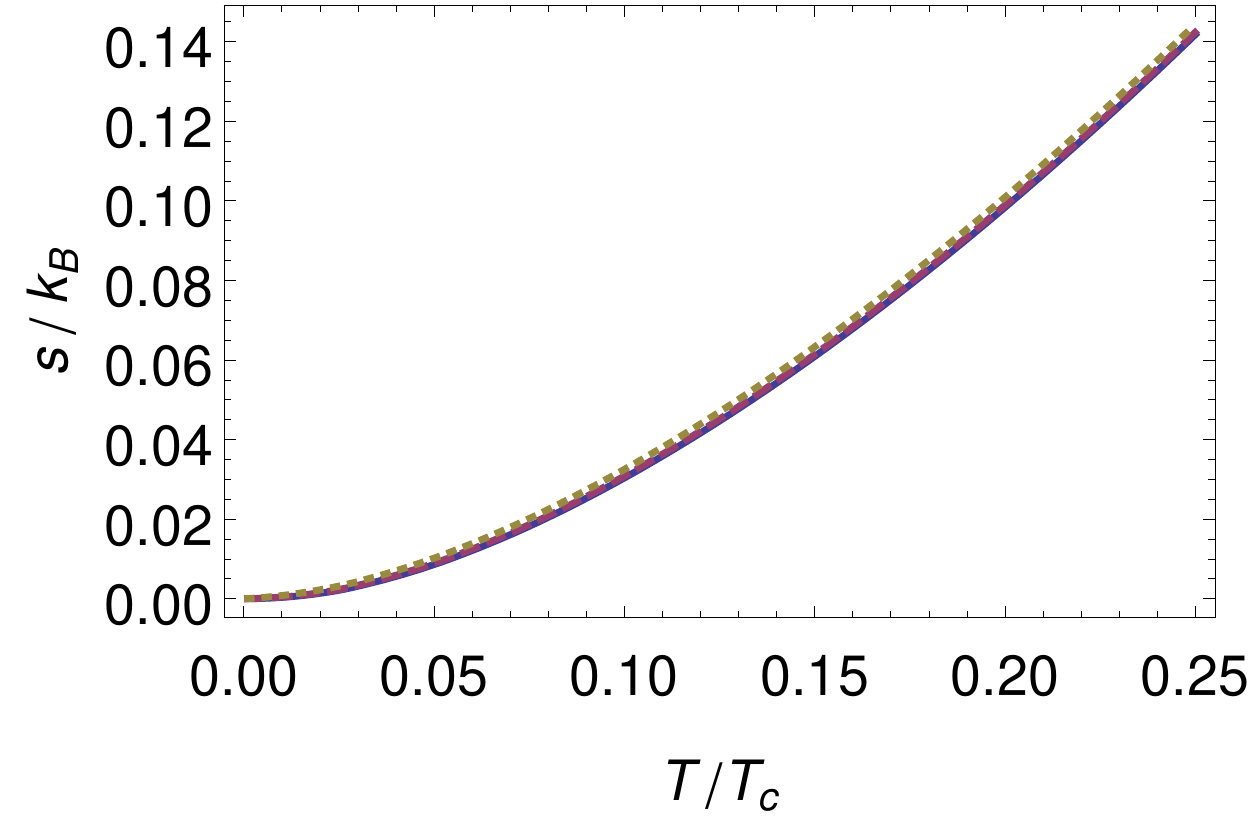}
\includegraphics[width=.3\textwidth]{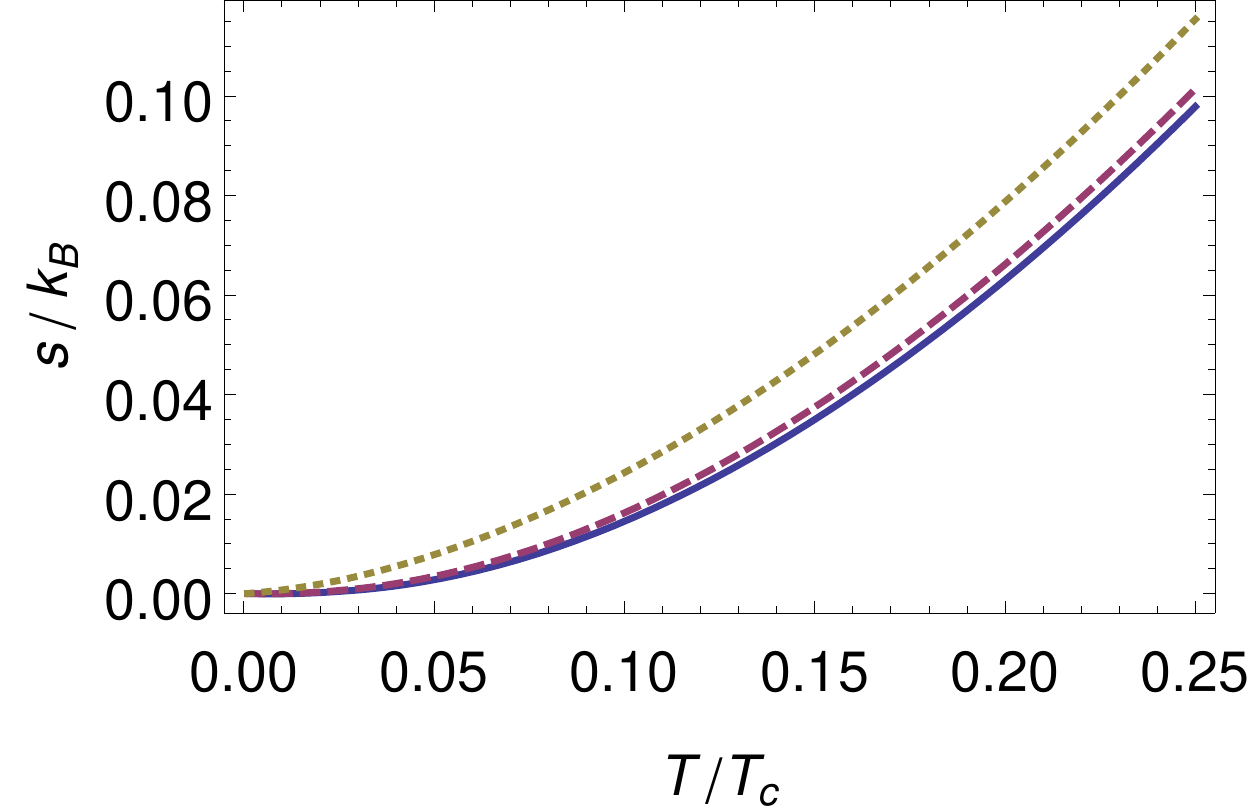}
\includegraphics[width=.3\textwidth]{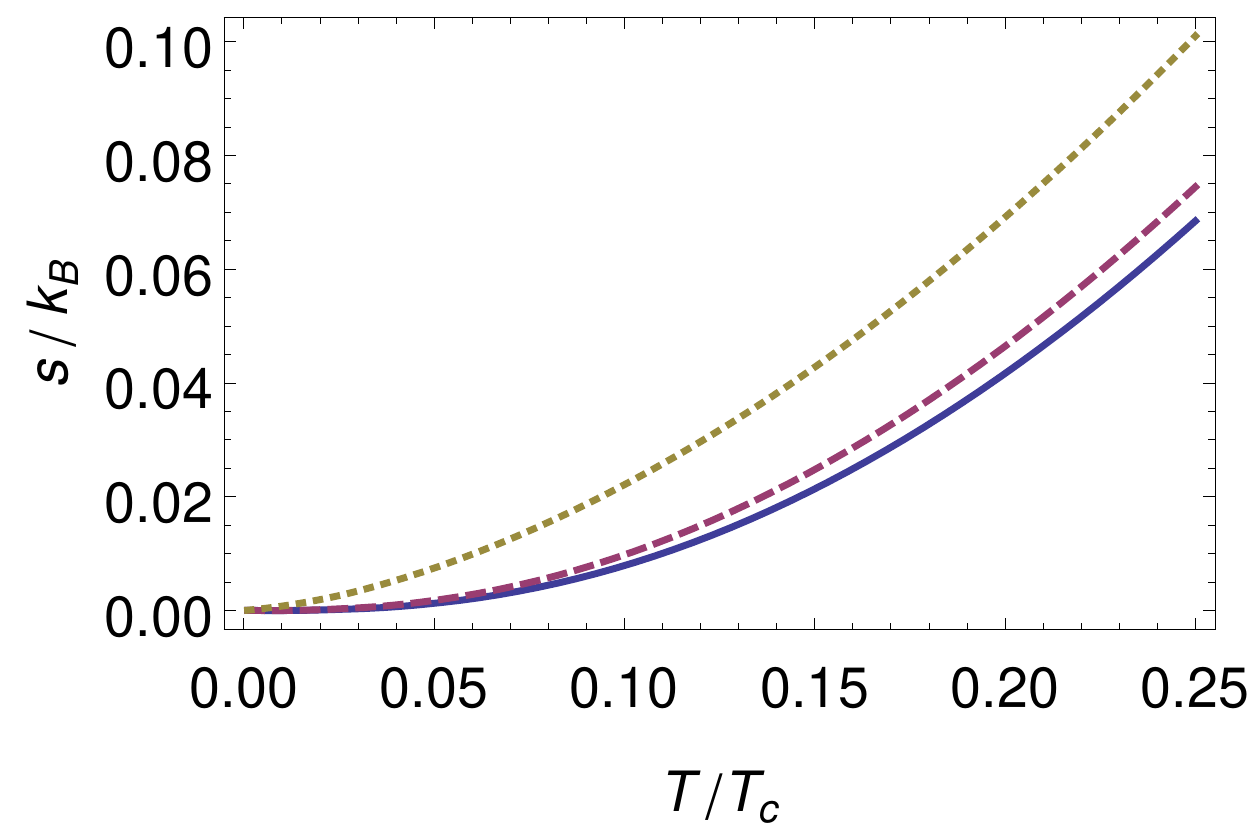}
\caption{(Color online) The entropy per particle of the two-species balanced gas.
The interaction parameter increases from left to right: $n^{1/3} a = 0.01$, $0.05$, and $0.1$. Intercomponent interactions are
zero ($\gamma=0$) for the solid line, moderate ($\gamma=1/2$)
for the dashed line, and strong ($\gamma\lesssim 1$) for the dotted line. A limited temperature range is depicted to emphasize the
difference between the curves. }
\label{fig-entropy}
\end{figure*}

The entropy per particle is presented in Fig.\ \ref{fig-entropy}.
Since the system is described by a mixture of noninteracting
phonon gases at low temperatures, the entropy per particle 
shows a $T^3$ power-law behavior at low temperatures. Furthermore,
the phonon velocity scales with the interaction strength and therefore
stronger-interacting gases have a higher entropy.

\section{Spin drag in a partially-condensed Bose gas}

\begin{figure*}[htp]
\centering
\subfigure[\ $C_{22}^{\ua\da}$]{
               \includegraphics[width=0.15\textwidth]{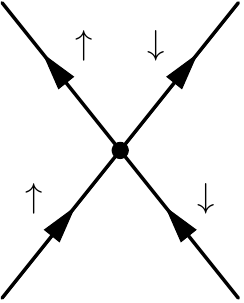}
               }
\subfigure[\ $C_{12}^{\ua\ua}$]{
               \includegraphics[width=0.15\textwidth]{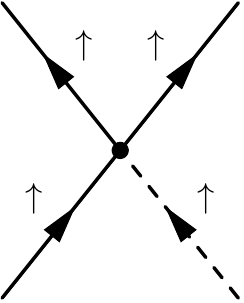}
               }
\subfigure[\ $C_{12}^{\da\da}$]{
               \includegraphics[width=0.15\textwidth]{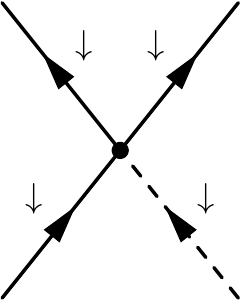}
               }
\subfigure[\ $C_{12}^{\ua\da}$]{
               \includegraphics[width=0.15\textwidth]{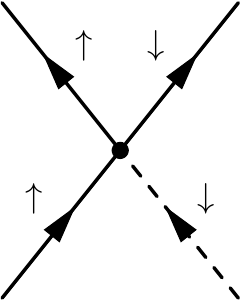}
               }
\subfigure[\ $C_{12}^{\da\ua}$]{
               \includegraphics[width=0.15\textwidth]{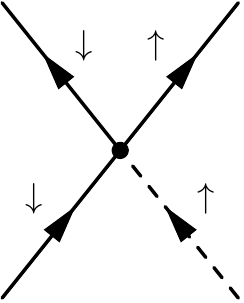}
               }
\caption{\label{fig:colints}Diagrams corresponding to the relevant collision integrals. Solid lines
        depict thermal atoms, while dashed lines represent condensed atoms.}
\end{figure*}

In this section we investigate kinetic (incoherent) processes that contribute to
density and velocity dynamics for the 
balanced binary Bose gas below the
critical temperature. 
We start by generalizing the results of the kinetic theory for a uniform single-species Bose
gas (see, for instance, Ref.\ \cite{ZNGuniform}) to the two-species gas. The collision integrals
that we consider below can be formally derived from the Heisenberg 
equations of motion for
the atoms of the two different species. For
a detailed discussion of that derivation for the case of a single 
species, we refer to Ref.\ \cite{ZNG}. The collision integrals
discussed in this section conserve the number of $\ua$ and
$\da$ particles separately.

Since the kinetic processes in question
vanish for very low temperatures, and in order to avoid
complications posed by the fact that the Bogoliubov transformation
mixes the spin species, we work in the Hartree-Fock approximation
in this section. The Hartree-Fock approximation is therefore
valid for the whole range of temperatures where the collision
processes play a significant role.
Moreover, here we consider the ``local 
equilibrium" situation where the chemical potentials
of the thermal atoms and the condensates are equal 
to $\mu = g_+ n + g n'$. The thermal
atoms thus feel the Hartree-Fock mean-field energy 
$(g_+ + g) n$. 
These considerations lead to a
Bose-Einstein distribution with a nonzero average momentum
$\bs p_{\text{nc}\ua}$ ($\bs p_{\text{nc}\da}$)
for the noncondensed $\ua$ ($\da$) atoms,
\be
f^\alpha_i = 
\frac{1}{
e^{\beta (\bs p_i-\bs p_{\text{nc}\sigma})^2/2m + \beta (g_+ + g) n -\beta\mu}-1
},
\ee
where $\sigma$ labels the spin species and $i$ labels the 
momentum variable $\bs p_i$ (the omission of the $i$ label denotes
the momentum variable $\bs p$). Furthermore, we allow for 
different nonzero momenta of the two condensates:
The $\ua$ ($\da$) condensate has a momentum
of $\bs p_{\text{c}\ua}$ ($\bs p_{\text{c}\da}$).

In the single-species case only one collision process,
namely $C_{12}$,
using the notation of Ref.\ \cite{ZNG},
is responsible for the relaxation of the difference between the
condensate velocity and the velocity of the thermal atoms. 
However, five different processes (see Fig.\ \ref{fig:colints})
can contribute to the dynamics of various densities and velocities
in the two-species system, leading to the following set of quantum
kinetic equations for the distributions of the thermal particles, 
where we only consider the collisional contributions:
\ba
\partial_t f^\uparrow |_\text{coll}
&=
C_{22}^{\ua\da}
+C_{12}^{\uparrow\uparrow}
+C_{12}^{\uparrow\downarrow}
+\bar C_{12}^{\downarrow\uparrow}
\label{eq:boltz1}
,
\\
%%%%%%%%%%%%%%%%%%%%%%%%%%%%%%%%%%%%%%%%%%
\partial_t f^\downarrow |_\text{coll}
&=
\bar C_{22}^{\ua\da}
+C_{12}^{\downarrow\downarrow}
+\bar C_{12}^{\uparrow\downarrow}
+C_{12}^{\downarrow\uparrow}
\label{eq:boltz2}
.
\ea

We now discuss the individual collision terms present in these
equations. We use a shorthand notation for the momentum
integrals
$\int_i \equiv \int d^3\bs p_i$ in order to simplify the
following formulas.
First, we have a 
spin-drag term 
(Fig.\ \ref{fig:colints} (a); cf.\ 
Refs.\ \cite{spindragTh, spindragExp}),
which only
involves scattering between the thermal atoms and thus
exists both below and above the critical temperature,
%%%%%%%%%%%%%%%%%%%%%%%%%%%%%%%%%%%%%%%%%%%%%%%%%%%%%%%%%%%%%%%%%%%
\ba
C_{22}^{\ua\da}=
\int_{1234} A_{22} \big[
\delta^{(3)}(\bs p - \bs p_4)
-\delta^{(3)}(\bs p - \bs p_1)
\big],
\\
\bar C_{22}^{\ua\da}
= 
\int_{1234} A_{22} \big[
\delta^{(3)}(\bs p - \bs p_3)
-\delta^{(3)}(\bs p - \bs p_2)
\big],
\ea
where we have denoted the common part of the integrand by
\ba
A_{22}
=
\frac{g_{\uparrow\downarrow}^2}
{(2\pi)^5\hbar^7}
&\delta^{(3)}(\bs p_1 + \bs p_2 - \bs p_3 - \bs p_4)
\\
\times&\delta([\bs p_1^2 + \bs p_2^2 - \bs p_3^2 - \bs p_4^2]/2m)
\nonumber \\ \nonumber
\times
\big[
f_1^\ua f_2^\da(1+&f_3^\da)(1+f_4^\ua)
-
(1+f_1^\ua)(1+ f_2^\da)f_3^\da f_4^\ua
\big].
\ea
This term contributes to the relaxation of the
difference of momenta $\bs p_{\text{nc}\ua}-\bs p_{\text{nc}\da}$.
Furthermore, we consider the following intraspecies
collision terms [Figs.\ \ref{fig:colints}(b) and \ref{fig:colints}(c)], which explicitly depend on the condensate
density and thus only exist below the critical temperature,
%%%%%%%%%%%%%%%%%%%%%%%%%%%%%%%%%%%%%%%%%%%%%%%%%%%%%%%%%%%%%%%%%%%
\ba
C_{12}^{\ua\ua(\da\da)}
=
&\int_{123} A_{12}^{\ua\ua(\da\da)} 
\big[
\delta^{(3)}(\bs p - \bs p_1)
\nonumber \\
&-\delta^{(3)}(\bs p - \bs p_2)
-\delta^{(3)}(\bs p - \bs p_3)
\big],
\ea
where
\ba
A_{12}^{\ua\ua(\da\da)}
=&
\frac{2 g^2 n_{0\ua(\da)}}
{(2\pi)^2\hbar^4}
\delta^{(3)}(\bs p_1 + \bs p_{\text c \ua(\da)} - \bs p_2 - \bs p_3)
\\
\times \delta([\bs p_1^2 &+ \bs p_{\text c \ua(\da)}^2 - \bs p_2^2 - \bs p_3^2]/2m -gn_{0\ua(\da)})
\nonumber \\ \nonumber
\times
\big[
f_1^{\ua(\da)} &(1+f_2^{\ua(\da)})(1+f_3^{\ua(\da)})
-
(1+f_1^{\ua(\da)})f_2^{\ua(\da)} f_3^{\ua(\da)}
\big].
\ea
These terms contribute to the dynamics of the condensate fraction
and also describe the relaxation between the condensate velocity
and the average thermal particle velocity of the same species.
Moreover, we also have similar terms for the interspecies
scattering, namely, 
\ba
C_{12}^{\ua\da(\da\ua)}
&=
\int_{123} A_{12}^{\ua\da(\da\ua)} \big[
\delta^{(3)}(\bs p - \bs p_2)
-\delta^{(3)}(\bs p - \bs p_1)
\big],
\\
\bar C_{12}^{\ua\da(\da\ua)}
&=
\int_{123} A_{12}^{\ua\da(\da\ua)}
\delta^{(3)}(\bs p - \bs p_3),
\ea
where
\ba
A_{12}^{\ua\da(\ua\da)}
&=
\frac{g_{\uparrow\downarrow}^2 n_{0\da(\ua)}}
{(2\pi)^2\hbar^4}
\delta^{(3)}(\bs p_1 + \bs p_{\text c \da(\ua)} - \bs p_2 - \bs p_3)
\\
\times
\delta([\bs p_1^2 &+ \bs p_{\text c \da(\ua)}^2 - \bs p_2^2 - \bs p_3^2]/2m
-gn_{0\da(\ua)}
)
\nonumber \\ \nonumber
\times
\big[
f_1^{\ua(\da)} &(1+f_2^{\ua(\da)})(1+f_3^{\da(\ua)})
-
(1+f_1^{\ua(\da)})f_2^{\ua(\da)} f_3^{\da(\ua)}
\big].
\ea
The latter terms also describe the dynamics of the condensate fraction,
as well as the relaxation of various velocities mediated
by the condensate.

In order to obtain the equations for the change of the density and
the momentum of the thermal particles, we perform
an integration of Eqs.\ \eqref{eq:boltz1} and \eqref{eq:boltz2}
leading to
\ba
\partial_t n'_{\ua(\da)}=
\int \frac{d^3\bs p}{(2\pi\hbar)^3} \partial_t f_{\ua(\da)},
\\
\partial_t ( n'_{\ua(\da)} 
\bs p_{\text{nc}\ua(\da)}) =
\int \frac{d^3\bs p}{(2\pi\hbar)^3} \bs p \partial_t f_{\ua(\da)}.
\ea
Adding the equations that change the thermal densities
and the equations
that change the condensate densities
obtained from straightforward considerations of the
collision integrals, we have
\ba
\partial_t n'_\ua
&=
\Gamma_{12}^{\uparrow\uparrow}
+\bar \Gamma_{12}^{\downarrow\uparrow}
,
\\
%%%%%%%%%%%%%%%%%%%%%%%%%%%%%%%%%%%%%%%%%%
\partial_t n'_\da
&=
\Gamma_{12}^{\downarrow\downarrow}
+\bar \Gamma_{12}^{\uparrow\downarrow}
,
\\
%%%%%%%%%%%%%%%%%%%%%%%%%%%%%%%%%%%%%%%%%%
\partial_t n_{0\ua}
&=-\Gamma_{12}^{\uparrow\uparrow}
-\bar \Gamma_{12}^{\downarrow\uparrow},
\\
%%%%%%%%%%%%%%%%%%%%%%%%%%%%%%%%%%%%%%%%%%
\partial_t n_{0\da}
&=-\Gamma_{12}^{\downarrow\downarrow}
-\bar \Gamma_{12}^{\uparrow\downarrow},
\ea
where 
\ba
\Gamma_{12}^{\ua\ua(\da\da)} &= -\frac{1}{(2\pi\hbar)^3}
\int_{123} A_{12}^{\ua\ua(\da\da)},
\\
\bar \Gamma_{12}^{\ua\da(\da\ua)} &= \frac{1}{(2\pi\hbar)^3}
\int_{123} A_{12}^{\ua\da(\ua\da)},
\ea
and we see that the total densities $n_{\ua(\da)}$ are conserved separately.
In a similar manner, the equations for the change of momenta are
\ba
\partial_t ( n'_{\ua} \bs p_{\text{nc}\ua})
&=
\frac{1}{(2\pi\hbar)^3}
\int_{1234} (\bs p_4 - \bs p_1) A_{22}
- \bs p_{\text{c}\ua} \Gamma_{12}^{\ua\ua}
\nonumber\\
+
\frac{1}{(2\pi\hbar)^3}&
\int_{123} (\bs p_2 - \bs p_1) A_{12}^{\ua\da}
+
\frac{1}{(2\pi\hbar)^3}
\int_{123} \bs p_3 A_{12}^{\da\ua},
\\
\partial_t ( n'_{\da} \bs p_{\text{nc}\da})
&=
-
\frac{1}{(2\pi\hbar)^3}
\int_{1234} (\bs p_4 - \bs p_1) A_{22}
- \bs p_{\text{c}\da} \Gamma_{12}^{\da\da}
\nonumber \\
+
\frac{1}{(2\pi\hbar)^3}&
\int_{123} \bs p_3 A_{12}^{\ua\da}
+
\frac{1}{(2\pi\hbar)^3}
\int_{123} (\bs p_2 - \bs p_1) A_{12}^{\da\ua},
\\
\partial_t ( n_{0\ua} \bs p_{\text{c}\ua})
&=
\bs p_{\text{c}\ua} \Gamma_{12}^{\ua\ua}
+
\bs p_{\text{c}\ua}
\Gamma_{12}^{\da\ua},
\\
\partial_t ( n_{0\da} \bs p_{\text{c}\da})
&=
\bs p_{\text{c}\da} \Gamma_{12}^{\da\da}
+
\bs p_{\text{c}\da}
\Gamma_{12}^{\ua\da}.
\ea
Note that the total momentum is conserved; thus,
\be
\partial_t ( n'_{\ua} \bs p_{\text{nc}\ua} +
n'_{\da} \bs p_{\text{nc}\da}+
n_{0\ua} \bs p_{\text{c}\ua}+
n_{0\da} \bs p_{\text{c}\da}) = 0.
\ee

For the hydrodynamic theory in the subsequent section, we are 
interested in the linearization of the collision integrals
in terms of the velocity differences. It is straightforward
to show that all the $\Gamma_{12}^{\alpha\beta}$ integrals 
are at least
quadratic in terms of the momenta and, therefore, in linear
response the densities
stay constant for both species and 
$\partial_t n'_\alpha = \partial_t n_{0\alpha} = 0$.
In addition, this result implies that the condensate momentum
experiences no linear relaxation. This is consistent with the 
common physical intuition that the condensate motion should not decay 
in the lowest order, as the condensate motion corresponds to the 
flow of a superfluid.

%1213-generating...
\begin{figure}[htp]
\centering
\includegraphics[width=.40\textwidth]{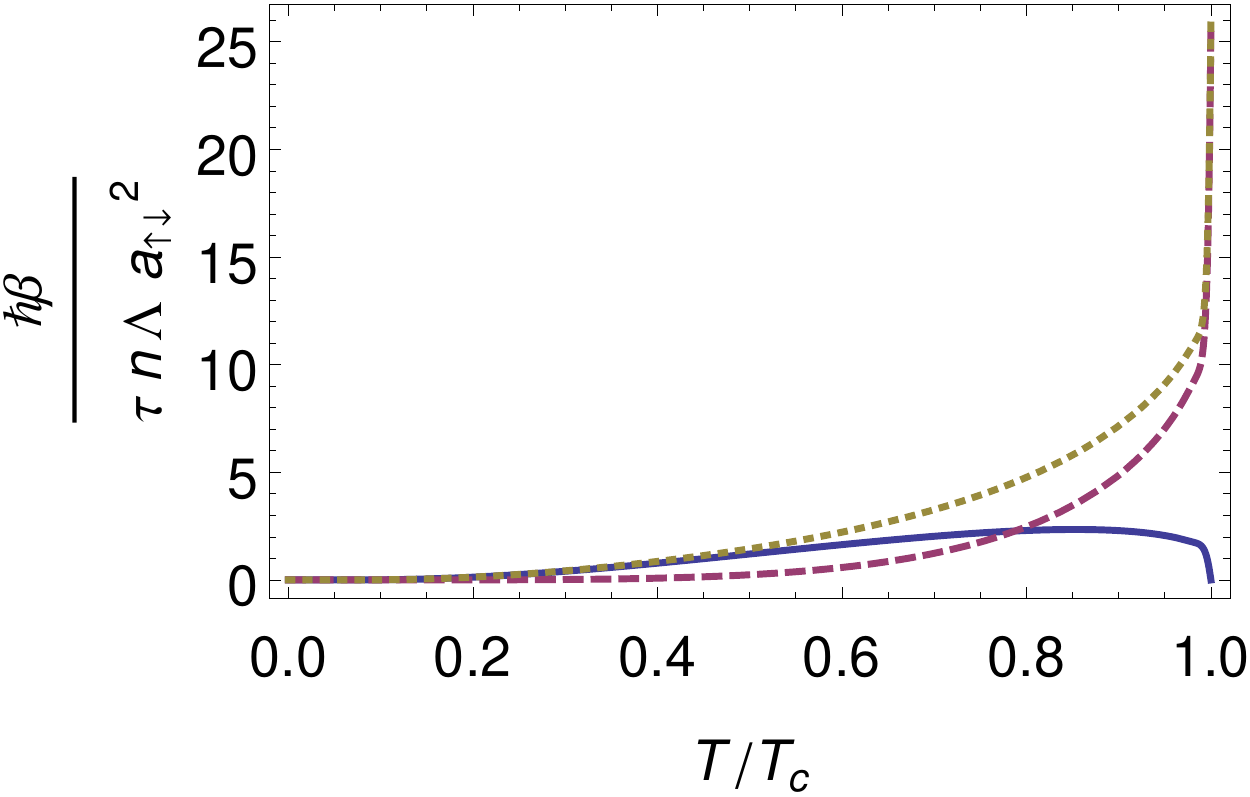}
\caption{(Color online) Condensate-assisted (solid line), thermal (dashed line), and total (dotted line)
spin-drag rates for a system of sodium atoms
at a density of $10^{21}$~m$^{-3}$, $\gamma\simeq1$, $n^{1/3}a = 0.03$.}
\label{fig-spindrag}
\end{figure}

Nevertheless, the momentum difference between the two
species of thermal particles experiences relaxation,
\ba
\partial_t (\bs p_{\text{nc}\ua} - \bs p_{\text{nc}\da})
&=
-\left(
\frac{1}{\tau_{22}}
+
\frac{1}{\tau_{12}}
\right)
(\bs p_{\text{nc}\ua} - \bs p_{\text{nc}\da})
\nonumber \\
&=
-
\frac{1}{\tau_\text{sd}}
(\bs p_{\text{nc}\ua} - \bs p_{\text{nc}\da}),
\ea
where we have also defined the total spin-drag
rate $1/\tau_\text{sd}$ in terms of the thermal
and condensate-assisted spin-drag relaxation rates $1/\tau_{22}$ and
$1/\tau_{12}$, respectively.
While the thermal spin-drag relaxation time
\ba
&\frac{\hbar\beta}{\tau_{22}n\Lambda a_{\ua\da}^2}
=
\frac{1}{6 \pi^2}
\frac{1}{(n \Lambda^3)^2}
\int_0^\infty
\frac{dq \, d\omega\,
q^2}{\sinh^2(\omega/2)}
\\ \nonumber
&
\phantom{=}
\times \ln\left(
\frac{e^{q^2/16\pi+\beta gn_{0}-\omega/2+\pi\omega^2/q^2}-e^{-\omega}}
{e^{q^2/16\pi+\beta gn_{0}-\omega/2+\pi\omega^2/q^2}-1}
\right)^2
\ea
has been calculated before in Ref.\ \cite{HedwigSpinDrag}
for the temperatures above the condensation temperature,
the condensate-assisted spin-drag rate,
\ba
&\frac{\hbar\beta}{\tau_{12}n\Lambda a_{\ua\da}^2}
=
\frac{64}{3(2\pi)^3}\frac{n_0}{n}\frac{a}{\Lambda}
\int_0^\infty
d p_1\, d p_3\, p_1 \, p_3^3
\\ \nonumber
&\phantom{=}\times\left(
1+\frac{1}{e^{(p_1^2+p_3^2)/4\pi+2\beta g n_0}-1}
\right)
\nonumber \\ \nonumber
&\phantom{=}\times \frac{1}{e^{p_1^2/4\pi+\beta g n_0}-1}
\frac{1}{e^{p_3^2/4\pi+\beta g n_0}-1}
\Theta\left(\frac{p_1 p_3}{2\pi} - \beta g n_0 \right),
\ea
where $\Theta$ denotes the Heaviside step function,
has not been investigated before.
As we can expect, the thermal spin-drag rate $1/\tau_{22}$ dominates at high
temperatures, while the condensate-assisted spin-drag rate 
$1/\tau_{12}$ is more important at
low temperatures (see Fig.\ \ref{fig-spindrag}).
Even though in our approximation the thermal spin-drag rate
has a maximum at the critical temperature, a more careful
calculation leads to its suppression due to critical
fluctuations 
\cite{RakpongSpindrag}
in a very narrow temperature window around $T_c$.
Therefore, strictly speaking, $1/\tau_{12}$ and $1/\tau_{22}$
vanish at both $T = 0$ and $T = T_c$.

\section{Hydrodynamic theory for a balanced binary mixture}

The goal of this section is to derive a set of hydrodynamic equations
for a balanced mixture of two components, where each of the components
has a superfluid part and a normal part. Note that the
following discussion is, in principle, not limited to weakly interacting
Bose gases, since it is only based on conservation laws. However,
we do not discuss various dissipative terms such as 
the thermal diffusivity or the (second) viscosity 
\cite{KhalatBook, NGtrappedHydro01, PeterHeat, HoShenoy98}, with the exception of the
spin-drag term. Furthermore, in the previous section we have concluded that the spin-drag term
only relaxes the noncondensate spin currents, whereas in what follows we discuss the hydrodynamics
in terms of superfluid and nonsuperfluid (normal) currents. Therefore, we posit that the nonsuperfluid
spin currents relax in exactly the same manner as the noncondensate spin currents. This is an approximation,
which is only valid in weakly interacting systems, where the condensate and the superfluid
are very similar objects.

In the equilibrium situation, the particle densities of the two 
components are identical for both the superfluid 
($n^\tsf_\ua = n^\tsf_\da\equiv n^\tsf$) and the normal fluid ($n^\tnf_\ua = n^\tnf_\da\equiv n^\tnf$), and therefore the total densities of each species are identical,
too ($n_\ua = n^\tsf_\ua + n^\tnf_\ua = n_\da$).
Moreover, in equilibrium there are no particle currents 
$\bs j^\tnf_\ua = \bs j^\tnf_\da = \bs j^\tsf_\ua = \bs j^\tsf_\da$, 
as all the velocities vanish: $\bs v^\tnf_\ua = \bs v^\tnf_\da = \bs v^\tsf_\ua = \bs v^\tsf_\da$, where the normal current of the
$\ua$ component is defined as $\bs j^\tnf_\ua = n^\tnf_\ua \bs v^\tnf_\ua$, and the other currents are defined similarly.
The current of each of the components is the sum of the
superfluid and normal currents: $\bs j_\ua = \bs j^\tsf_\ua + \bs j^\tnf_\ua$.

In the nonequilibrium situation, however, both the velocities
and the various densities can be nonzero and different from
each other. 
In that case, it is useful to define variables pertaining to the 
combined motion of the whole gas and contrast them to the
variables that describe the relative motion. Therefore,
\ba
\bs j^{\tnf,\tsf}_\tot = \bs j^{\tnf,\tsf}_\ua + \bs j^{\tnf,\tsf}_\da,
\ea
are the total normal and superfluid currents of the combined
motion, respectively. Adding these two currents up yields the
total particle current $\bs j_\tot = \bs j^\tnf_\tot + \bs j^\tsf_\tot$.
Furthermore, we are now in a position to define the combined
velocities such that
\be
\bs j^\tnf_\tot \equiv
n^\tnf_\tot \bs v_\tot^\tnf,
\ee
and
\be
\bs j^\tsf_\tot \equiv
n^\tsf_\tot \bs v_\tot^\tsf.
\ee

Note also that thermodynamic functions (e.g., pressure and entropy)
are defined for the whole system only, and not for the individual
components. Moreover, due to time-reversal invariance,
the lowest-order velocity corrections to thermodynamic functions are
quadratic \cite{KhalatBook} and therefore do not enter linear
equations. For example, the Gibbs-Duhem relation reads
\be
n_\ua d\mu_\ua + n_\da d\mu_\da + s dT = dp
\ee
even in the case of nonzero velocities. Hence, we obtain 
the following linearized hydrodynamic equations
for the combined (or in-phase) motion of the whole gas:
\ba
\partial_t n_\tot + \bs \nabla \cdot \bs j_\tot = 0, \\
\partial_t (n_\tot s) 
+ n_\tot s \bs \nabla \cdot \bs v_\tot^\tnf = 0, \\
m\partial_t \bs j_\tot + \bs \nabla p = 0, \\
m\partial_t \bs v_\tot^\tsf + \bs \nabla \mu_\tot =0.
\ea
These simple equations express particle conservation, entropy conservation,
Newton's second law, and the Josephson relation, respectively. Note that the terms 
``in-phase'' and ``out-of-phase'' in this paper refer to the motion of the two spin components, as opposed
to the normal and the superfluid component, where the latter meaning is common in the literature concerning
liquid helium.

We now turn our attention to the relative motion of the up and down particles. To that end, we define the normal and superfluid
spin densities
\be
\Delta n^{\tsf,\tnf} = n_{\ua}^{\tsf,\tnf} - n_{\da}^{\tsf,\tnf},
\ee
as well as spin currents
\be
\Delta\bs{j}^{\tsf,\tnf}
=
\bs j^{\tnf,\tsf}_\ua - \bs j^{\tnf,\tsf}_\da.
\ee
Using these newly defined quantities, the hydrodynamic
spin equations are the following:
\ba
\partial_t \Delta n^\tnf + \bs \nabla \cdot \Delta\bs{j}^\tnf &=0,\\
\partial_t \Delta n^\tsf + \bs \nabla \cdot \Delta\bs{j}^\tsf &=0,\\
\partial_t \Delta\bs{j}^\tnf 
+n_\tot^\tnf \bs \nabla \Delta \mu /2m
&= -\Delta\bs{j}^\tnf /\tau_{sd}, \\
\partial_t \Delta\bs{j}^\tsf 
+n_\tot^\tsf \bs \nabla \Delta \mu /2m
&= 
0.
\ea
The four equations above
state that both normal and superfluid spin densities are conserved and
the normal spin current is driven by a chemical potential difference
and relaxes with the spin-drag rate, whereas the superfluid spin
current is also driven by a chemical potential difference but does
not relax.

By eliminating all the currents and velocities from the equations
above and noticing that for the balanced case
\be
\frac{\partial \Delta \mu}{\partial T}
=
\frac{\partial \Delta \mu}{\partial n}
=0,
\ee
we find that
\ba
\label{eq:excitations1}
m \partial_t^2 n_\tot &= \bs \nabla^2 p,\\
m \partial_t^2 s &= \frac{n^\tsf}{n^\tnf} s^2 \bs \nabla^2 T,\ea
\ba
\partial_t^2 \Delta n^\tnf + \frac{1}{\tau_{sd}}\partial_t 
\Delta n^\tnf
&=
 \\ \nonumber &\phantom{=}
\frac{n^\tnf}{m}\frac{\partial \Delta \mu}{\partial \Delta n}\bs \nabla^2
(\Delta n^\tnf + \Delta n^\tsf), \\
\partial_t^2 \Delta n^\tsf
&=
\label{eq:excitations4}
 \\ \nonumber
&\phantom{=}\frac{n^\tsf}{m}\frac{\partial \Delta \mu}{\partial \Delta n}\bs \nabla^2
(\Delta n^\tnf + \Delta n^\tsf).
\ea
We now employ a traveling-wave ansatz for
the total density
\be
n_\tot = n_{\tot,\text{eq}} + \delta n_\tot \exp(-i[\omega t-\bs k \cdot \bs x]),
\ee
where $n_{\tot,\text{eq}}$ is the total density in the equilibrium, and 
$\delta n_\tot$ is the amplitude of the wave. Proceeding similarly for the other quantities, 
we obtain the lowest-lying 
collective modes for the system. Note that in this
balanced situation the in-phase collective modes (first and second sounds) 
are decoupled from the out-of-phase modes (spin modes).

\section{Results}

%1205-TEST...
\begin{figure*}[htp]
\centering
\includegraphics[width=.3\textwidth]{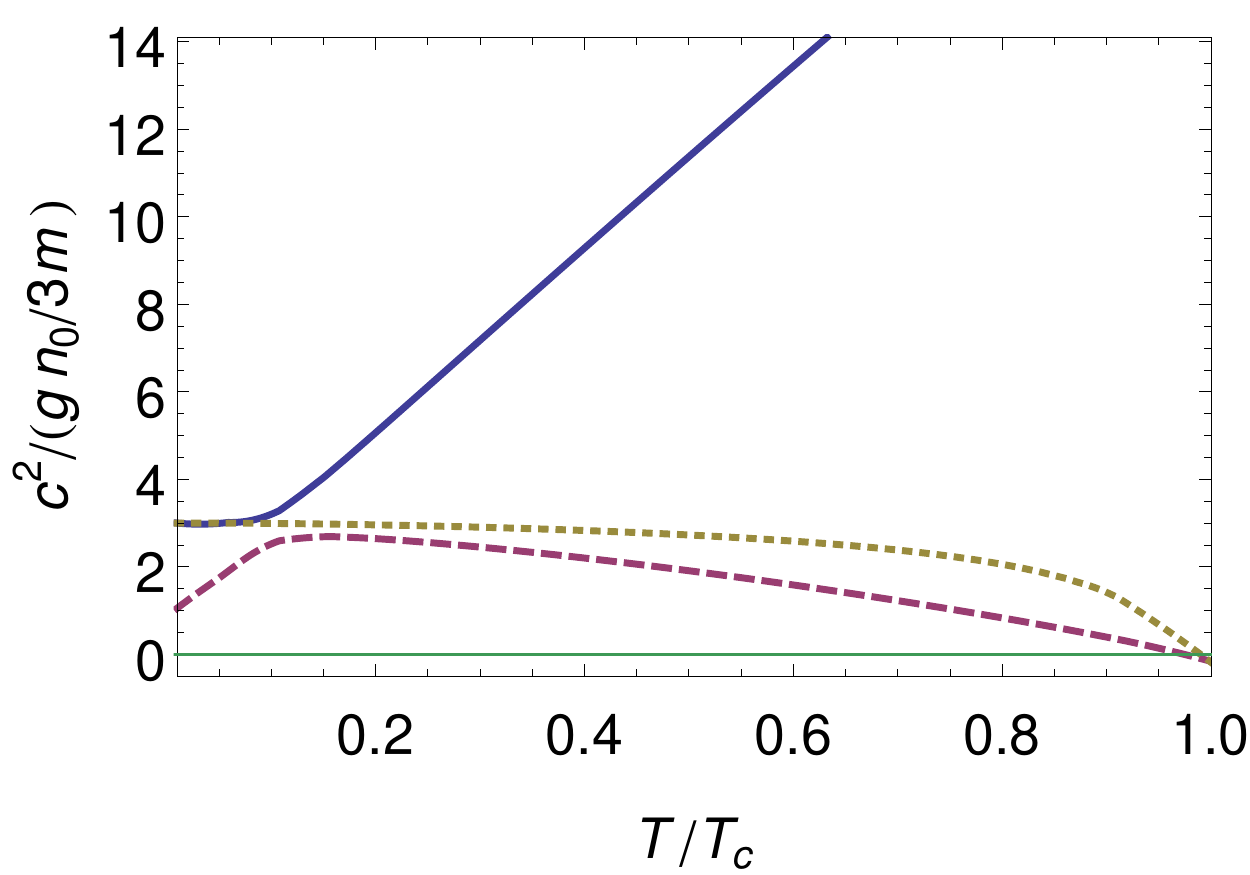}
\includegraphics[width=.3\textwidth]{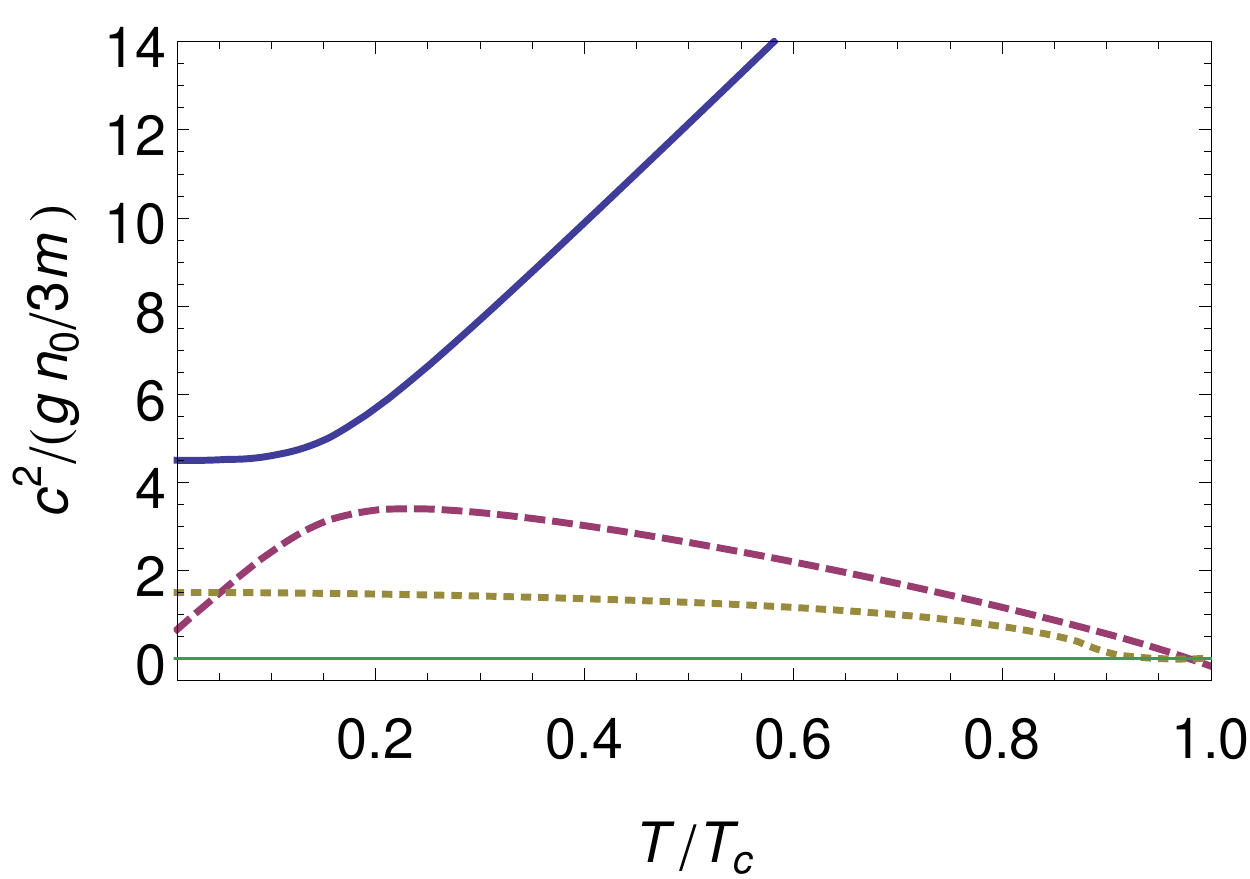}
\includegraphics[width=.3\textwidth]{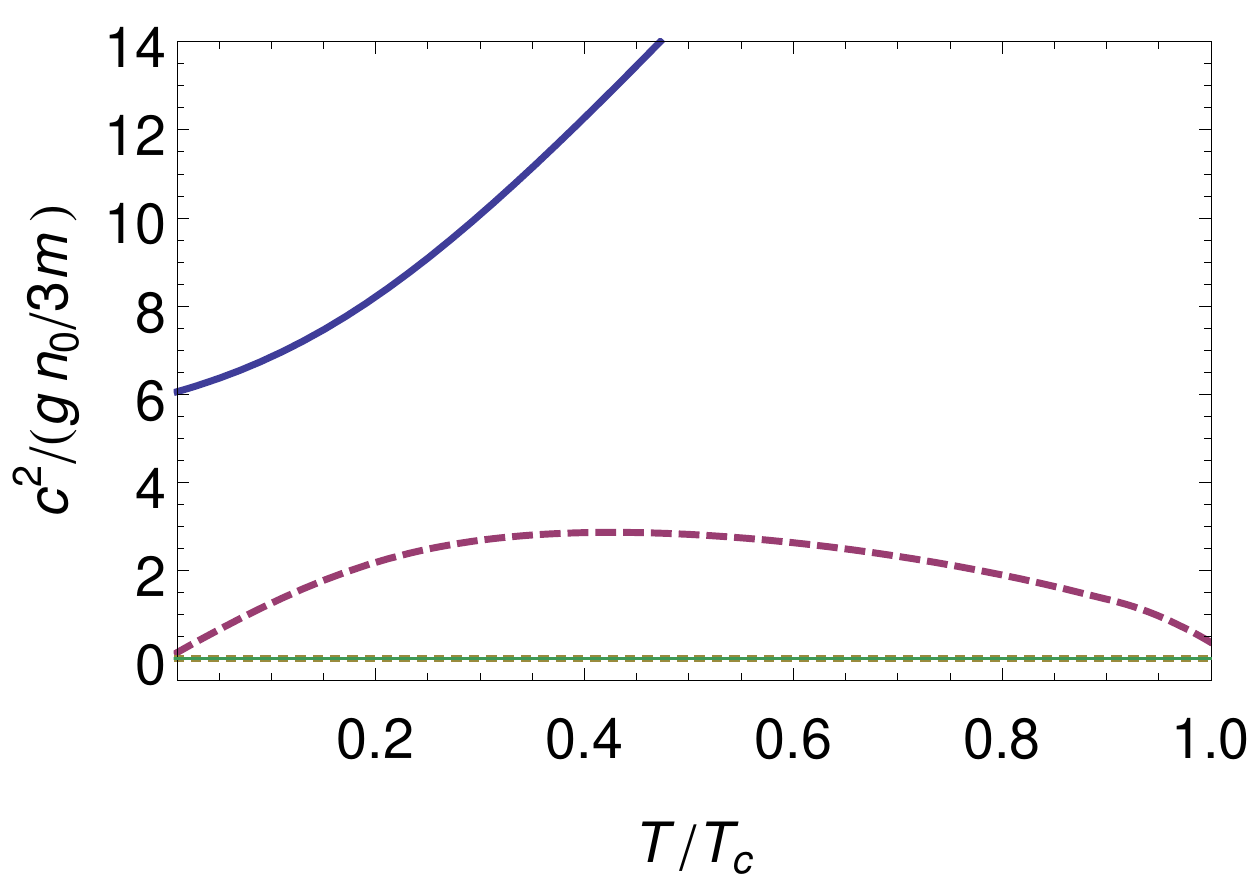}
\caption{(Color online) Square of the sound velocities of the two-species balanced 
Bose gas scaled to the zero-temperature second-sound velocity of a
single species. Here $n^{1/3} a = 0.03$. 
The inter-component interactions are
increasing from left to right: $\gamma = 0$ (left), $\gamma=1/2$ 
(center), and $\gamma\lesssim 1$ (right). The latter corresponds
to a mixture of hyperfine states of the sodium atom
with the density of $10^{21}$~m$^{-3}$. The first-sound 
velocity is depicted by the thick solid line, the second-sound
velocity is the dashed line. The spin-sound velocities
are the dotted and the thin solid lines.  }
\label{fig-sounds2}
\end{figure*}

In this section we present the results of our calculations, most important of which are the sound velocities
of the various modes. We discuss both the uniform and the trapped cases. From the total number of hydrodynamic
equations, we expect four modes: two in-phase modes and two out-of-phase modes. The two in-phase
modes are present in any system where entropy transport is distinct from density transport: They are the 
first- and the second-sound modes. It is worthwhile to note that the phenomenon of the second sound 
has been reported not only in superfluid helium, but in other systems as well, including solid helium 
\cite{2ndSoundSolidHe}, dielectric crystals \cite{2ndSoundNaF}, and, more recently, a unitary Fermi gas 
\cite{Fermi2ndSound13}. Moreover, contrary to the case of superfluid helium, second sound in weakly 
interacting gases is not a pure entropy wave, as discussed below. 
On the other hand, the out-of-phase modes are similar to spin modes
and  therefore only occur in systems with several species of particles.

\subsection{Collective modes in a uniform gas}
In the case of no spin drag, all the collective modes of the system
have a linear dispersion
\be
\omega = c_i k,
\ee
where $c_i$ are the sound velocities.
At zero temperature, the first-sound velocity can be calculated 
from the expression of pressure in Eq.\ \eqref{eq:pressure}, yielding
\be
c_1^2 = \frac{g_+ n}{m} = \frac{(1+\gamma) gn}{m}.
\ee
Moreover, the second-sound velocity can be calculated in a manner
similar to that used in the single-component case (see Ref.\ \cite{PandSbook}
for an explicit calculation), the only difference being that,
instead of a single phonon gas, we have to consider a mixture
of two noninteracting phonon gases, leading to
\be
c_2^2 = \frac{gn}{3m}
\frac{(1+\gamma)(1-\gamma)^{5/2}+(1-\gamma)(1+\gamma)^{5/2}}
{(1-\gamma)^{5/2}+(1+\gamma)^{5/2}}.
\ee
When the two components are decoupled ($\gamma=0$), 
we recover the single-species result $c_2^2 = gn/3m$ from the formula above. 
Furthermore, in the strong-coupling
limit ($\gamma \lesssim 1$), the sound velocity vanishes ($c_2^2 \simeq 0$),
signaling the demixing transition.
When it comes to the spin sounds, one of them always has a zero velocity, whereas
the other one has the velocity
\be
c_s^2 = \frac{g_- n}{m} = \frac{(1-\gamma)g n}{m},
\label{eq:cs2}
\ee
as can be seen from the difference of chemical potentials
in Eq.\ \eqref{eq:chempots}. Note that $c_1$ and $c_s$ can also
be obtained by expanding the energies of the Bogoliubov excitations
in Eq.\ \eqref{eq:balbog} to the lowest order in momentum. Therefore,
the first sound and the spin sound can be thought of as both
quasiparticles and collective modes at zero temperature.

\begin{figure*}[htp]
\centering
\includegraphics[width=.3\textwidth]{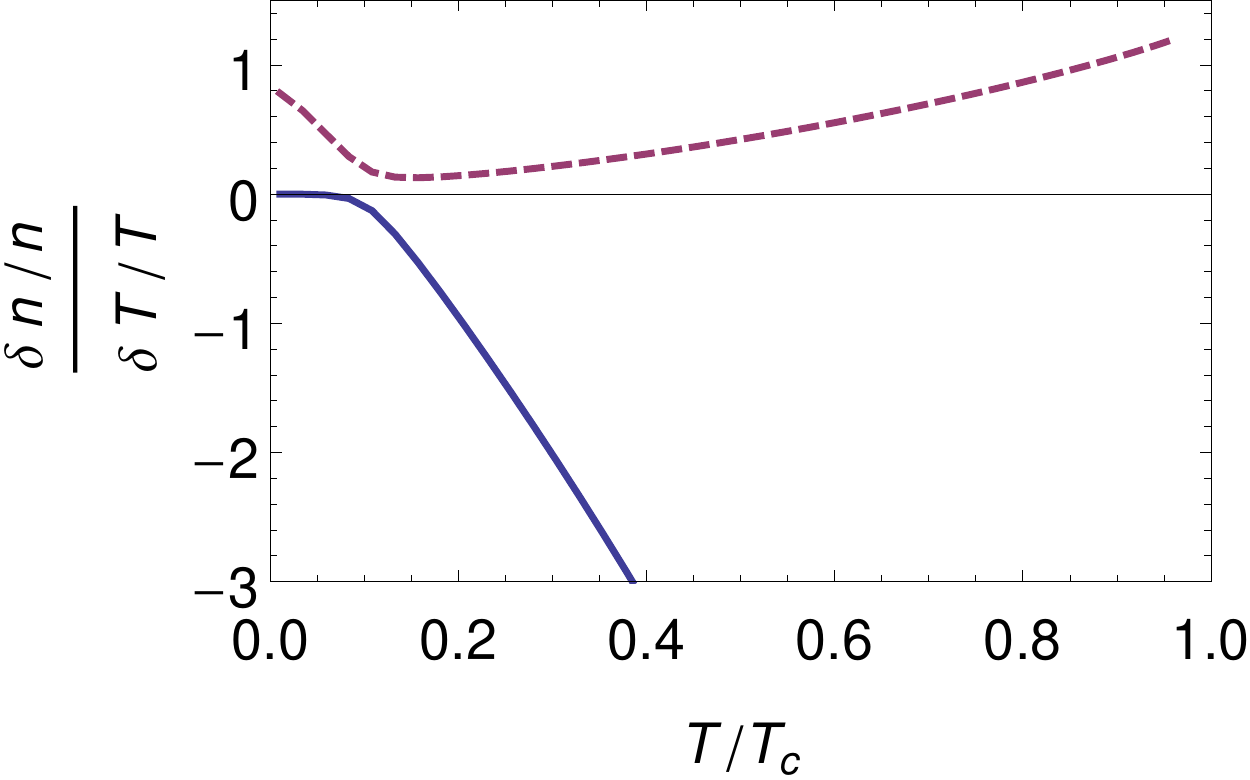}
\includegraphics[width=.3\textwidth]{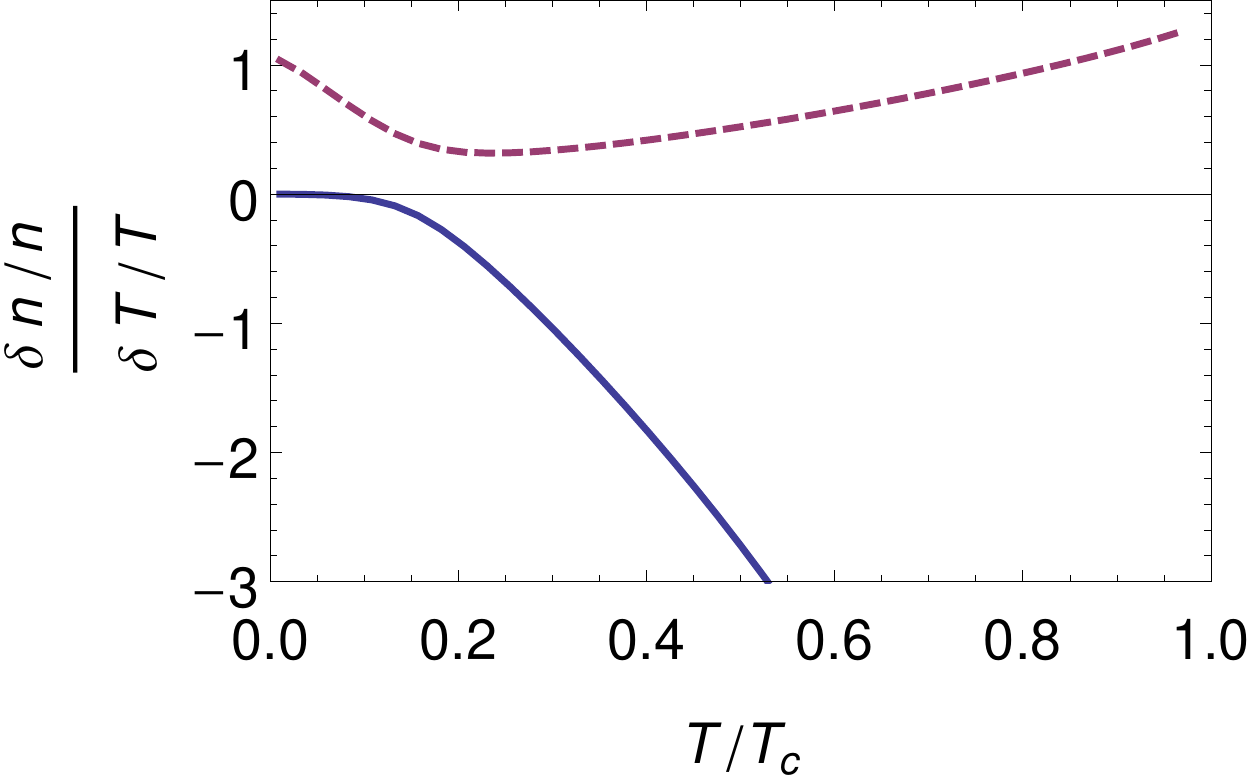}
\includegraphics[width=.3\textwidth]{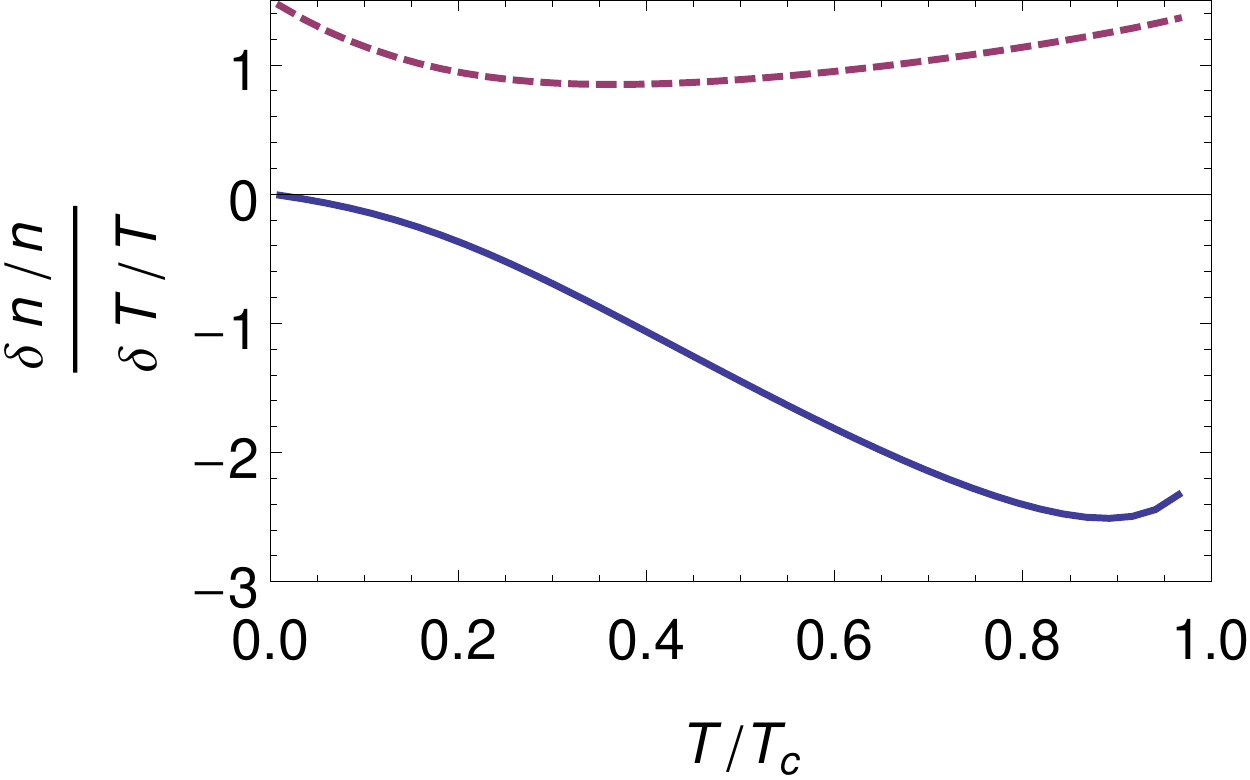}
\caption{(Color online) Ratio of density and temperature fluctuations present
in first sound (solid line) and second sound (dashed line)
for $n^{1/3} a = 0.03$. 
Intercomponent interactions are
increasing from left to right: $\gamma = 0$ (left), $\gamma=1/2$ 
(center), and $\gamma\lesssim 1$ (right). The latter corresponds
to a mixture of hyperfine states of the sodium atom
with the density of $10^{21}$~m$^{-3}$.}
\label{fig-soundchar2}
\end{figure*}

We present the
velocities of the sound modes for different temperatures in Fig.\ \ref{fig-sounds2}. In order to characterize the sound modes, 
we have calculated the amplitudes of the temperature and density perturbations 
from the eigenvectors of the linearized system (see
Fig.\ \ref{fig-soundchar2}). We observe that for the temperatures
below the avoided crossing, first sound is mostly a temperature
wave. However, above the avoided crossing temperature, first sound 
is predominantly a density wave. Second sound has comparable 
contributions from both relative temperature and density deviations
for any temperature. Furthermore, second sound is a wave where
density and temperature change in phase, while first sound
describes an out-of-phase change (the temperature increases with 
decreasing density), as signified by a minus sign in Fig.\ 
\ref{fig-soundchar2}. When it comes to the spin modes, 
a zero-frequency mode exists, which corresponds to
\be
\Delta n^\tnf = -\Delta n^\tsf,
\ee
and does not affect the total density of either component.
The other spin mode, however, affects
the total spin density, while the normal component contributes with the same
relative weight as the superfluid component:
\be
\frac{\Delta n^\tnf}{n^\tnf} = \frac{\Delta n^\tsf}{n^\tsf}.
\label{eq:spin-density-mode}
\ee

\begin{figure}[htp]
\centering
\includegraphics[width=.40\textwidth]{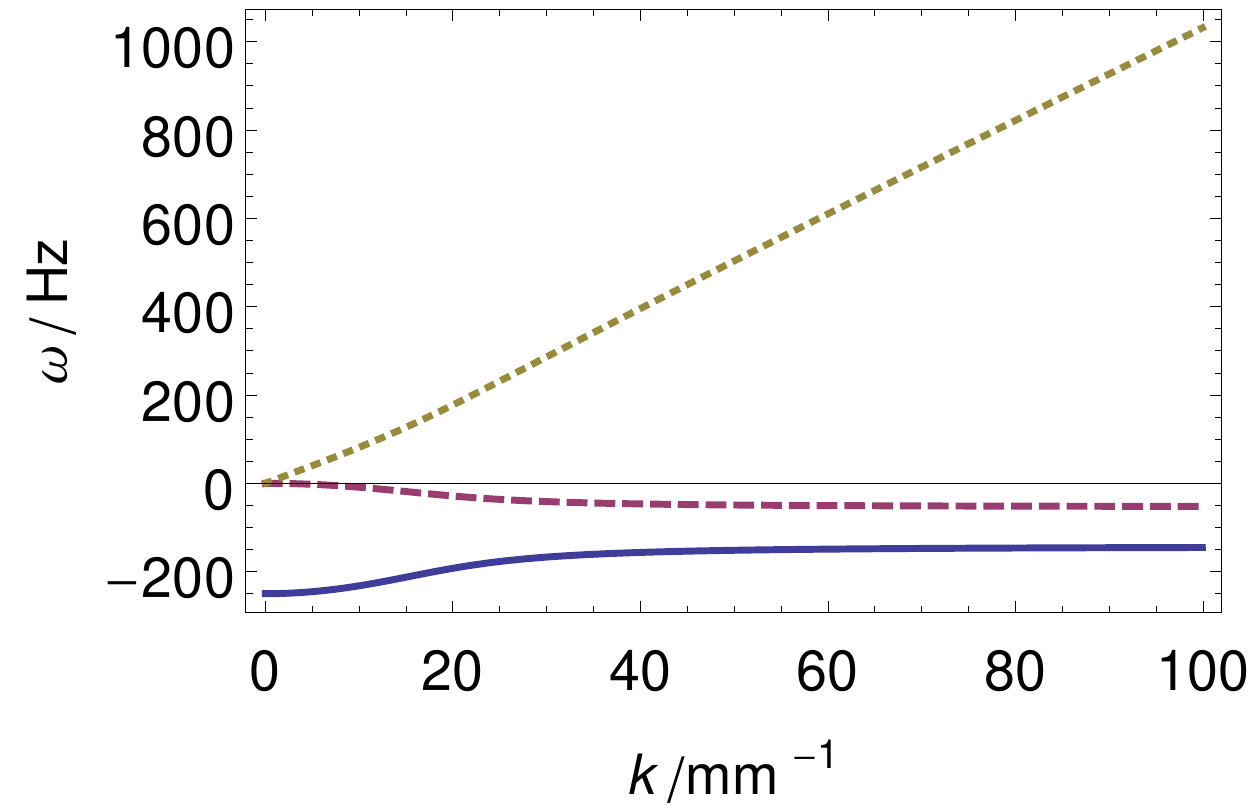}
\caption{(Color online) 
Spin modes in the $n^{1/3} a = 0.03$ case at $T=T_c/2$ for nonzero interspecies interactions 
($\gamma=1/2$, $1/\tau_{sd} = 0.25$kHz). The solid
line depicts the imaginary mode, whereas the dashed and
dotted lines depict the imaginary and the real parts of
the complex spin mode. The horizontal axis corresponds to
the physically relevant wavelengths down to about $10\mu\text{m}$.
 }
\label{fig-spindrag-modes}
\end{figure}

Upon accounting for spin drag, we find that the zero-frequency modes split
into two: one zero-frequency mode and one purely imaginary
mode. The purely imaginary mode (cf.\ Fig.\ \ref{fig-spindrag-modes})
at low momenta is an excitation with only normal density fluctuations, whereas
at higher momenta it preserves the total density of every component
as the zero-frequency mode: $\Delta n^\tnf = -\Delta n^\tsf$.
Furthermore, the dispersion of the other spin
mode develops a quadratic imaginary part, even though its real
velocity $c_s$ is almost unaffected by spin drag and the mode is still characterized
by Eq.\ \eqref{eq:spin-density-mode}. The frequency $\omega$ of the latter
mode at long wavelengths can be written as
\be
\omega = c_s k - i D k^2,
\ee
where $k$ is the wavenumber and $D$ is the diffusion coefficient. Hence, we call this mode the complex spin mode in Fig.\ \ref{fig-spindrag-modes}.

Since all the interaction strengths are similar ($\gamma \simeq 1$) in a sodium gas,
when spin drag is absent, both spin modes have zero frequency.
In particular, the spin sound has zero velocity at any temperature, $c_s \simeq 0$,
which is consistent with the zero-temperature result in Eq.\ \eqref{eq:cs2}.
When spin drag is present, the purely imaginary mode has a
constant imaginary frequency $\omega = -i/\tau_{sd}$, while the complex mode
frequency remains zero, and no imaginary part develops, so $D\simeq 0$ in this case.
In a realistic
sodium gas with a density of $10^{21}$ m$^{-3}$ at half of the critical
temperature, the total spin-drag rate is 
$1/\tau_\text{sd}\simeq 1$~kHz, while in experiments of trapped 
sodium gas \cite{spindragExp} above $T_c$ typical rates are on the 
order of $0.1$~kHz. Therefore, it should be experimentally possible to
measure the spin-drag rate below the critical temperature by
measuring the spin-wave decay time.

%1205-TEST...

\subsection{Gas in a trap}

\begin{figure}[htp]
\centering
\includegraphics[width=.40\textwidth]{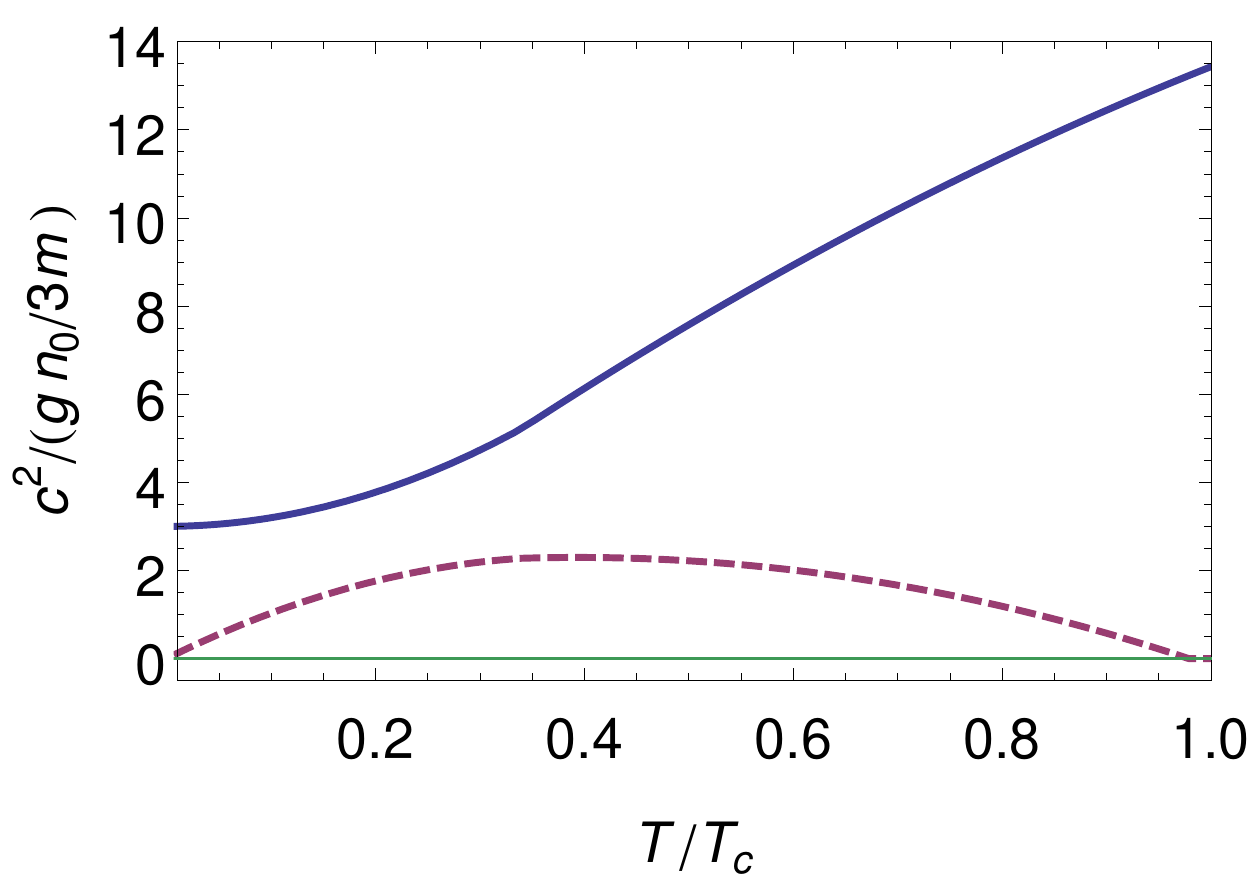}
\caption{(Color online) Trap averaged sound velocities normalized to the peak density in the trap with no damping (see text for details).}
\label{fig-trap-avg-sounds}
\end{figure}

\begin{figure}[htp]
\centering
\includegraphics[width=.40\textwidth]{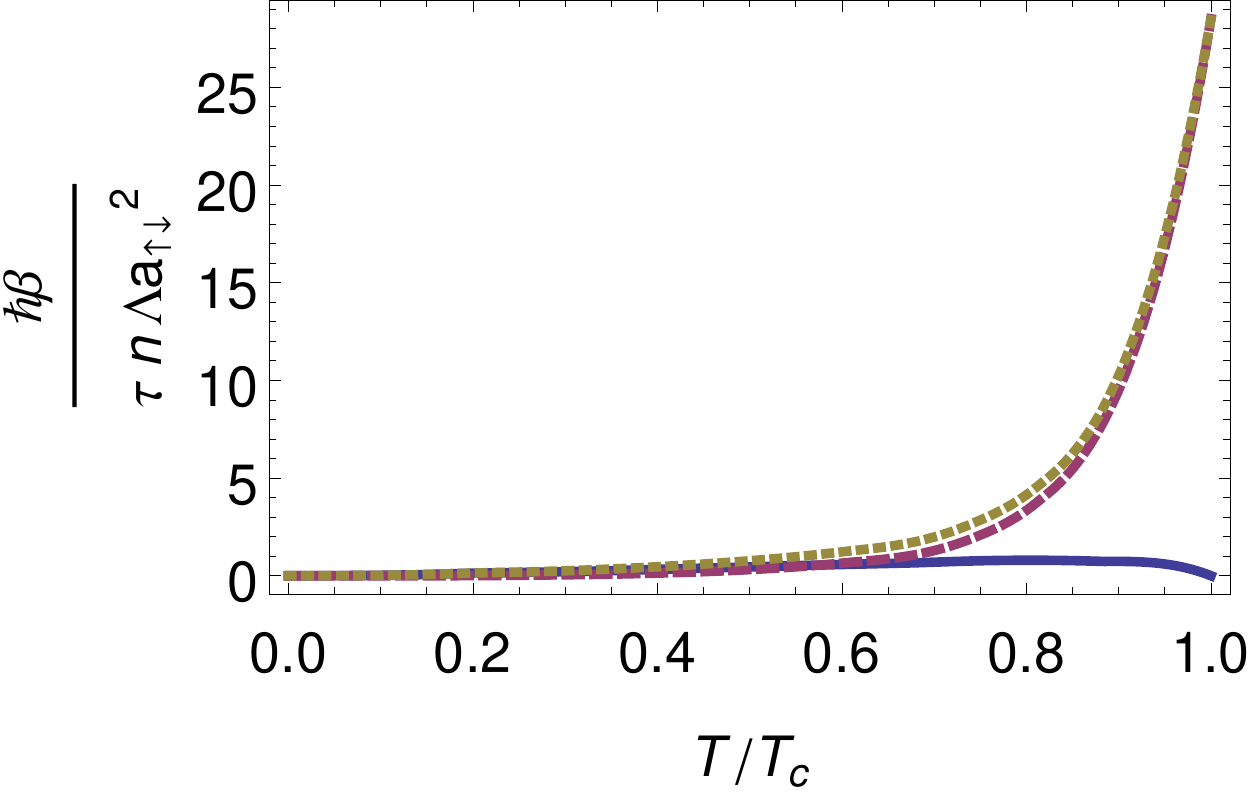}
\caption{(Color online) 
Trap averaged  condensate-assisted (solid line), thermal (dashed line), and total (dotted line)
spin-drag rates normalized to the peak density in the trap (see text for details).}
\label{fig-trap-avg-spindrag}
\end{figure}

We now turn our attention to the experimentally relevant
trapped case. Here we consider a cylindrically symmetric trap which is highly
anisotropic. This trap, where one (axial) direction is
very shallow, and the other two (radial) directions are much more
strongly confined, puts the gas in the hydrodynamic regime in the
axial direction even in the presence of weak interactions. Therefore, we perform the
trap average in the radial direction and analyze the propagation of excitations in the axial direction.
To that end, we work in the semi-classical approximation (see Ref. \cite{PandSbook} for more details).
The condensate is treated in the Thomas-Fermi approximation and 
the excitations are treated in the Hartree-Fock approximation. Therefore, both components
are treated in the local-density approximation.
We consider the experimentally relevant situation \cite{PeterHeat} of
$2\cdot 10^9$ sodium atoms
in a trap with an axial trapping frequency of $2\pi \times 2$ Hz and a radial trapping frequency of
$2\pi \times 100$ Hz, and extract the radial profile at the center of the trap from this calculation.
We then perform an average on all the thermodynamic quantities of the hydrodynamic Eqs.\
\eqref{eq:excitations1}--\eqref{eq:excitations4}. The resulting sound velocities are presented in 
Fig.\ \ref{fig-trap-avg-sounds}. At zero temperature, the first-sound velocity in a trap is suppressed by a factor
of $\sqrt{2}$ as compared to the uniform case, since the average density in the trap is half of the peak density in the Thomas-Fermi approximation, i.e.,
\be
\langle n_0 \rangle = n_0/2,
\ee
as first pointed out by Zaremba \cite{ZarembaTrap}. Since at zero temperature the second-sound velocity
is also proportional to the square root of the density, it is suppressed by the same factor of $\sqrt{2}$. 
This is not explicit for sodium atoms as the second sound at zero temperature vanishes.
However, for a nonzero temperature we have not succeeded in finding a similar simple relation between
the trapped and uniform gases. Finally, we also present the spin-drag rate dependence on temperature
for a trapped system (Fig.\ \ref{fig-trap-avg-spindrag}). The curves are qualitatively similar to the uniform
case, even though the rates are decreased by a factor of about $10$.

%1221-trap-average...

%1221-trap-average-spindrag.nb

\section{Conclusion}

In summary, we have constructed a hydrodynamic theory of a balanced two-species Bose mixture.
We have also calculated the microscopic thermodynamic parameters entering that theory using the Popov approximation,
obtaining the equation of state on the way. Moreover, we have accounted for the relaxation of the
normal current by the spin-drag mechanism, considering the condensate-mediated spin-drag term
for the first time. Adding these components together we were able to calculate the sound velocities
and spin-drag rates for the experimentally accessible system of sodium atoms in a trap. We hope that
our analysis will stimulate experimental work on hydrodynamic modes and spin drag in a partially condensed Bose mixture.

\acknowledgments
This work was supported by the Stichting voor Fundamenteel Onderzoek der Materie (FOM) and the European Research Council (ERC) and is part of the D-ITP consortium, a program of the Netherlands Organisation for Scientific Research (NWO) that is funded by the Dutch Ministry of Education, Culture and Science (OCW).
It is our pleasure to thank M. Di Liberto, S. Grubinskas, and C. Khripkov for stimulating discussions.

\bibliography{bogoliubov}

\begin{thebibliography}{70}
\bibitem [{ome()}]{omega}%
  \BibitemOpen
  \href@noop {} {}\bibinfo {note} {In the grand-canonical ensemble
  thermodynamic quantities are obtained by considering the grand potential
  $\oo$ that explicitly depends on the chemical potentials. In that case, the
  density is $n_\sigma = -\partial \oo / \partial \mu_\sigma |_T$, and the
  entropy per volume is $S/V = -\partial \oo / \partial T | _{\mu_\sigma}$. We
  have checked that the quantities obtained that way match the ones given in
  the main text. The said calculation was not presented here in detail, since
  the relevant expressions (including the dispersions of the quasiparticles)
  become rather cumbersome.}\BibitemShut {Stop}%
\end{thebibliography}

\end{document}